\documentclass[smallcondensed]{svjour3}     
\smartqed  

\input{0_Preamble}
\journalname{Empirical Software Engineering}
\begin{document}
\setcounter{tocdepth}{2}
\setcounter{secnumdepth}{3}

\title{Empirical Characterization of Logging Smells in Machine Learning Code}

\author{Patrick Loic Foalem \and Leuson Da Silva \and Foutse Khomh \and Heng Li \and Ettore Merlo
}

\authorrunning{Patrick Loic Foalem\and Leuson Da Silva \and Foutse Khomh \and Heng Li \and Ettore Merlo}


\institute{ \affmark[*]Corresponding author. \\
\\
           Patrick Loic Foalem \and  Leuson Da Silva \and Foutse Khomh \and Heng Li \and  Ettore Merlo \at
              \affaddr{Department of Computer Engineering and Software Engineering \\ Polytechnique Montreal \\
              Montreal, QC, Canada} \\
              \email{\{patrick-loic.foalem \and  leuson-mario-pedro.da-silva \and  foutse.khomh \and  heng.li \and ettore.merlo\}@polymtl.ca}           
}

\date{Received: date / Accepted: date}

\maketitle

\section*{Abstract}
Logging plays a central role in ensuring reproducibility, observability, and reliability in machine learning (ML) systems. While logging is generally considered a good engineering practice, poorly designed logging can negatively affect experiment tracking, security, debugging, and system performance. This study aims to create a taxonomy of logging smells specific to ML projects and to assess their prevalence and practical impact. 

We conducted a large-scale empirical analysis of 444 ML repositories and manually labeled 2,448 instances of logging smells. Based on this analysis, we constructed a taxonomy consisting of 12 ML-specific logging smell types spanning security, metric management, configuration, verbosity, and context-related issues. Our results show that logging smells are widespread in ML systems and vary in frequency and manifestation across projects.

To validate the taxonomy and understand its practical relevance, we conducted an online survey with 27 ML practitioners. The majority of respondents agreed with the defined smells and reported that several types—particularly Logging Sensitive Data, Metric Overwrite, Missing Hyperparameter Logging, and Log Without Context—have significant impact on reproducibility, maintainability, and system trustworthiness. At the same time, some smells (e.g., Heavy Data Logging and Print-based Logging) were perceived as context-dependent, reflecting trade-offs between performance and observability.

We publicly release our manually labeled dataset to support future research. We discuss implications for researchers, practitioners, and educators, and outline directions for automated detection and repair of logging smells using AI techniques. Our findings position logging quality as a critical and underexplored dimension of ML system engineering.


\section{Introduction}
\label{sec:introduction}
Logging is a fundamental practice in software engineering, essential for capturing information about the execution of software systems \citep{li2017log, foalem2024studying, chen2017characterizing}. It plays a critical role in various activities, including debugging \citep{li2020qualitative}, monitoring \citep{foalem2024studying}, and auditing \citep{foalem2025logging}, by providing valuable insights into system behavior during development and in production environments. Consequently, establishing and adhering to good logging practices is paramount for the long-term maintainability and comprehensibility of any complex system \citep{kernighan1999practice}. However, determining what information to log, where to insert logging statements, and at which verbosity level remains a non-trivial challenge for developers \citep{li2020qualitative}.

Code smells, a concept introduced by \citet{fowler2018refactoring}, refer to design or implementation flaws that, while not necessarily bugs, indicate potential issues that may impair software maintainability and scalability \citep{kaur2020systematic}. The presence of code smells and related anti-patterns has been shown to negatively impact software quality \citep{kaur2020systematic}, leading to increased maintenance effort, reduced comprehensibility, and a higher propensity for faults \citep{kaur2020systematic, cairo2018impact, abidi2021multi}. Therefore, the systematic study of code smells is driven by the need to identify and refactor these problematic areas to enhance software quality and reduce technical debt \citep{cedrim2017understanding, bibiano2019quantitative, lahti2021experiences}. Logging smells are particularly important to study because they affect the ability to monitor and audit software systems effectively \citep{kaur2020systematic, saarimaki2024taxonomy}. The identification of logging smells can aid in improving software observability, making it easier to track system behavior and ensure reliability.

Several studies have examined code smells in traditional software systems, focusing on identifying, cataloging, and analyzing the impact of various smells in object-oriented systems, particularly those written in languages like Java and C\# \citep{yamashita2013developers, palomba2014they, sharma2017house}. \citet{jabrayilzade2024taxonomy} propose a taxonomy of smells in inline code comments, while \citet{saarimaki2024taxonomy} introduced a taxonomy of code smells for traditional software. These studies have provided valuable insights into common issues and bad practices that can degrade software quality. By characterizing these smells, these studies have helped improve the overall quality of traditional software systems. Such research has also led to the development of automated tools for detecting and refactoring code smells, thereby supporting better software practices in the development lifecycle for traditional applications \citep{oztas2025towards, fawad2025refactoring}.

In recent years, Machine Learning models have transitioned from isolated experimental environments to being increasingly deployed in complex, real-world applications and integrated into traditional software systems \citep{dilhara2021understanding, amershi2019software, bosch2021engineering}. This integration necessitates effective and systematic logging, not only for the surrounding application but for the ML pipeline itself. Recent studies have begun to investigate logging practices within these ML-based systems \citep{foalem2024studying, foalem2025logging, rodriguez2025automated, chen2025empirical}. For instance, \citet{foalem2024studying} identified the use of at least 12 different logging libraries, which they broadly categorized into two main types: general-purpose libraries (e.g., logging \footnote{\url{https://docs.python.org/3/library/logging.html}}) and ML-specific libraries (e.g., MLflow \footnote{\url{https://mlflow.org/}}, TensorBoard\footnote{\url{https://www.tensorflow.org/tensorboard?hl=fr}}). This work highlights that logging in ML systems serves unique roles beyond traditional debugging, such as experiment tracking, data versioning, and model management. \citet{van2021prevalence} further demonstrate that ML projects exhibit widespread code smells distinct from those in traditional software. Their large-scale static analysis of 74 Python ML projects found pervasive issues such as code duplication and poor dependency management.
Despite emerging interest, existing ML smell taxonomies focus on general ML-specific code smells (e.g., reproducibility, dependency management, or misuse of ML APIs) and do not examine logging smells, even though logging is central to experiment tracking, governance, and model lifecycle management. Likewise, prior logging smell taxonomies are derived exclusively from traditional systems and overlook ML-specific logging failures.

This study aims to address this critical gap by conducting the first empirical characterization of logging smells specifically within Machine Learning code. Building upon the dataset of ML logging practices provided by \citet{foalem2024studying}, this paper investigates the following two research questions:

\begin{itemize}
    \item \textbf{RQ1: What logging smells exist in open-source ML-based systems?}
    \item \textbf{RQ2: How do ML practitioners perceive and experience these logging smells in practice?}
\end{itemize}

To answer \textit{RQ1}, we systematically analyzed 2,448 function-level logging instances collected from 444 open-source ML repositories. Through an iterative, human-in-the-loop taxonomy construction process supported by LLM-assisted classification, we derived a data-driven catalog of 12 recurring logging smells. 

To address \textit{RQ2}, we designed a structured practitioner survey aimed at validating the taxonomy and assessing its practical relevance. Following established methodologies in software engineering research, the survey collects background information, experience with ML logging tools, and practitioner evaluations of each smell type based on relevance, frequency, and perceived severity. The analysis of practitioner feedback shows that the proposed logging smell taxonomy largely aligns with real-world ML development practices, with strong agreement on its validity and high perceived relevance and severity for smells that threaten security, reproducibility, and experimental correctness, while also revealing context-dependent nuances for smells related to verbosity, configuration, and tooling ecosystems.

To summarize, the main contributions of this paper are: 

\begin{enumerate}
    \item We propose the first taxonomy of logging smells in ML code based on the evidence from 444 ML systems.
    \item We gather practitioners' insights about logging smells through a survey.
    \item We make our replication package \footnote{\url{https://github.com/foalem/LoggingSmellMLCode}} publicly available, including a labeled dataset of logging smells, to enable the research community to replicate, validate, and extend our study.
\end{enumerate}

Our findings provide valuable insights for ML practitioners by raising awareness of 12 common logging smells and their impact on security, reproducibility, observability, and experiment management. Early identification of these issues can prevent costly downstream consequences such as misleading metrics, irreproducible results, or performance degradation. Moreover, by releasing a labeled dataset of 2,448 logging smell instances, we enable researchers to develop automated detection and AI-assisted repair techniques, advancing tooling support for improving logging quality in ML systems.

\textbf{Organization:} The rest of the paper is organized as follows: Section \ref{sec:related_work} presents background and related work on logging and code smell from different perspectives, while Section \ref{sec:approach} explains the experimental setup to answer our research questions. The research question results are presented in Section \ref{sec:result}, followed by the implications of these results in Section \ref{sec:discussion_implication}. Section \ref{sec:threats_to_validity} discusses potential threats to the validity of the study. Finally, in Section \ref{sec:conclusion}, we present our conclusion and suggest avenues for future research.

\section{Background and Related Works}
\label{sec:related_work}
In this section, we discuss related work on logging and code smell in both traditional and machine learning software systems. 
\subsection{Logging in Traditional and ML-Based Systems}
Research on logging in traditional software systems has established a foundational understanding of why developers log, how they decide what to log, and the challenges they face. \citet{rong2017systematic} conducted a systematic review of logging practice and showed that developers frequently struggle to determine appropriate log levels, craft meaningful log messages, and maintain logging code as software evolves. Similarly, \citet{gu2022logging} mapped logging research across decades and highlighted recurring issues such as insufficient contextual information, inconsistent logging styles, and the difficulty of balancing verbosity with performance overhead. These works demonstrate that logging is not merely an auxiliary task; it is an essential development activity intertwined with debugging, maintenance, and program comprehension.

While traditional systems focus on control-flow and state--change logging, ML-based systems introduce additional complexity that these foundational studies do not address. \citet{foalem2024studying} showed that ML applications exhibit significantly sparser logging--approximately one log per 1{,}150 lines of code--despite requiring richer execution traces to understand data flows, model behavior, and training dynamics. Their study also revealed that developers face uncertainty about what to log in ML pipelines, particularly when dealing with hyperparameters, dataset statistics, intermediate metrics, and model checkpoints. This gap underscores a fundamental misalignment between traditional logging practices and the unique observability requirements of ML systems.

Further analysis of ML logging challenges by \citet{batoun2024literature} revealed that traditional logging mechanisms are insufficient for capturing the non-deterministic, data-dependent, and iterative nature of ML workflows. Their work demonstrated that ML debugging requires contextualizing logs with experiment metadata, model versions, random seeds, and performance curves—information that general-purpose logging libraries are not designed to manage. As a result, ML practitioners increasingly rely on specialized experiment-tracking tools.

The growing ecosystem of ML-specific logging and experiment tracking frameworks, such as MLflow, W\&B, Comet, Neptune, and TensorBoard, reflects this shift. \citet{rodriguez2025automated} examined automated log generation for ML applications using large language models (LLMs) and found that ML projects expose patterns and requirements that differ qualitatively from traditional systems. Their results showed that LLMs often misplace logs, overlog, or generate logs inconsistent with ML development norms, illustrating that ML-aware logging requires a deeper understanding of model training loops, evaluation procedures, and data transformations.

Taken together, these studies reveal a clear trajectory. Traditional logging research has provided invaluable insights into developer behavior, challenges, and best practices. However, ML-based systems extend the role of logging far beyond conventional debugging: logging must now serve experiment tracking, reproducibility, governance, and interpretability. Despite the emergence of ML-specific tools, empirical evidence indicates that logging in ML systems remains underdeveloped, inconsistent, and poorly understood. This gap motivates the need for a systematic examination of logging practice in ML codebases, particularly to identify and characterize the logging smells that arise uniquely in ML workflows.

\subsection{Code Smells in Traditional and ML-Based Systems}

Research on code smells has long focused on traditional software systems. Fowler’s early catalog established smells as recurring indicators of poor design and maintainability issues~\citep{fowler2018refactoring}, a perspective reinforced by empirical studies linking smells to increased defect-proneness and maintenance effort~\citep{cairo2018impact}. This body of work has motivated extensive research on automated smell detection and refactoring techniques~\citep{fontana2013code, schumacher2010building}, framing smells as actionable symptoms of structural degradation.

Within this broader landscape, logging-related issues have also received attention. \citet{saarimaki2024taxonomy} developed a comprehensive taxonomy of logging smells in traditional systems, highlighting problems such as inconsistent log levels, missing contextual information, and redundant messages. Complementary work by \citet{washizaki2019studying} showed how logging practices evolve over time and introduced recurring patterns of misuse, though their insights remain grounded in general-purpose, non-ML contexts. More recently, \citet{madi2025systematic} synthesized findings from 21 primary studies on logging smells, revealing fragmentation in terminology, datasets, and evaluation methods and underscoring the absence of a unified understanding of logging quality.

In contrast to these traditional software-level studies, work on code smells in machine learning systems is still emerging. A prominent effort by \citet{zhang2022code} systematically mined academic papers, grey literature, GitHub commits, and Stack Overflow posts to construct the first catalog of ML-specific code smells. Their catalog identifies 22 recurrent issues--ranging from misuse of ML library APIs to reproducibility hazards such as uncontrolled randomness--that stem from the unique characteristics of ML pipelines. Their findings emphasize that ML systems introduce new forms of technical debt, and that traditional smell taxonomies fail to capture ML-specific patterns related to data handling, stochastic training behavior, and framework-dependent APIs.

Together, these studies demonstrate that while traditional code smell research offers strong foundations, ML-based systems pose qualitatively different challenges due to dynamic training processes, heterogeneous frameworks (e.g., TensorFlow, PyTorch, Scikit-Learn), and the central role of experiment tracking and observability. Existing taxonomies do not account for the complex interactions between code, data, and model behavior that shape logging and monitoring practices in ML development.

Our work extends this emerging line of research by focusing specifically on logging smells in ML systems--an area not covered by prior ML smell catalogs--and provides the first empirical characterization of ML-specific logging pitfalls and their implications for traceability, reproducibility, and governance.


\section{Experiment setup} 
\label{sec:approach}
\begin{figure}[htp!]
    \centering
    \includegraphics[width=0.8\textwidth]{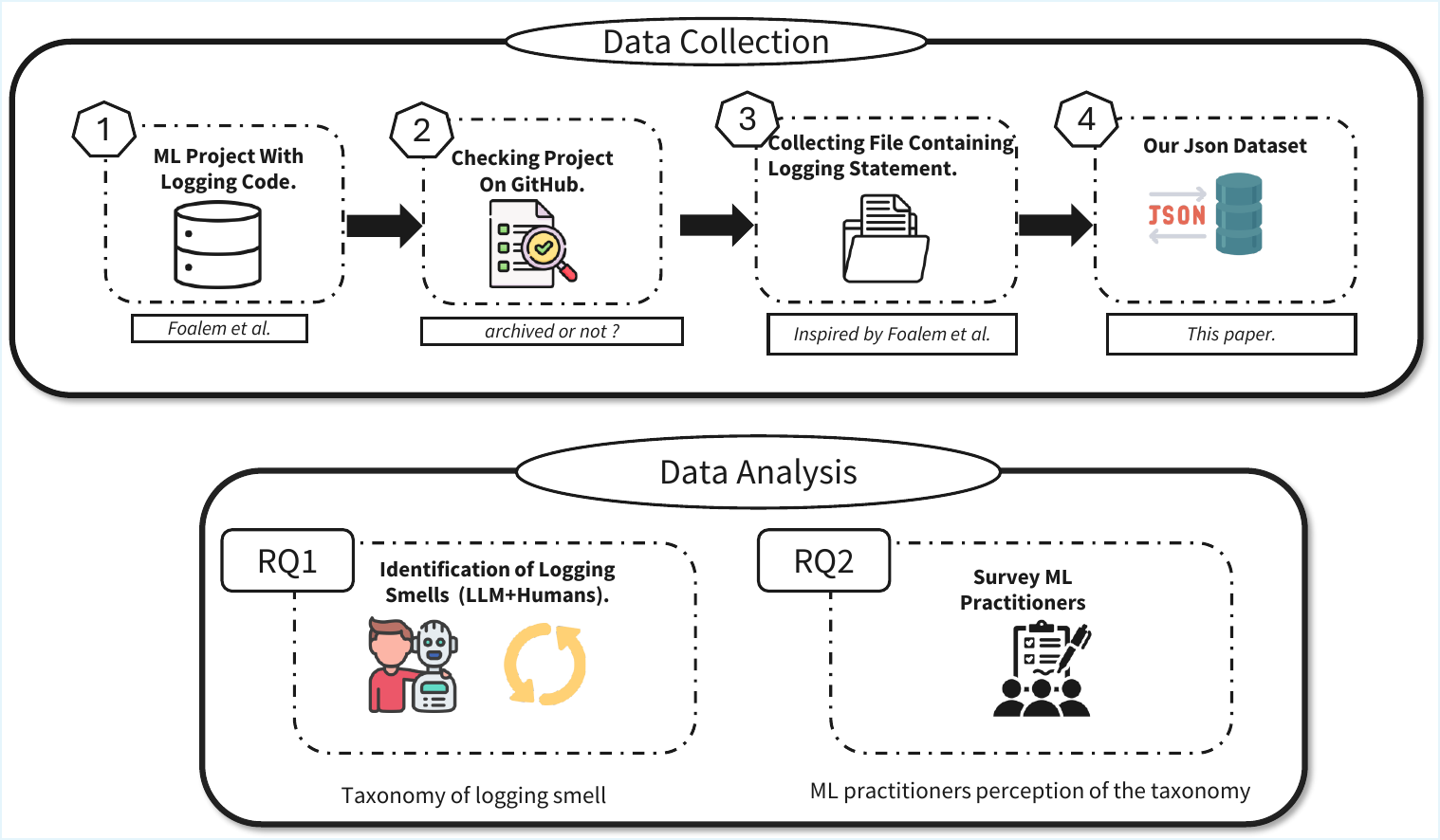}
    \caption{Overview of our research study.}
    \label{fig:researchprocess}
\end{figure}

We adopt a mixed-methods approach, combining both quantitative and qualitative methods \citep{ivankova2006using}. Specifically, we organized our methodology into two main steps, as presented in Figure~\ref{fig:researchprocess}. In the subsequent sections, we provide a detailed explanation of the first main step: data collection. Overall, this step focuses on collecting files containing logging statements inside ML code. After collecting these files, the second main step, data analysis, is presented in Section~\ref{sec:result}. There, we describe our qualitative analysis approach to answer our research questions RQ1 and RQ2 and report our results for each RQ.
\begin{enumerate}[label=\protect\circled{\arabic*},leftmargin=0.8 pt,align=left]
\item \textbf{ML project with logging code}: The first step of our data collection process involves adopting the dataset of ML logging practices provided by \citet{foalem2024studying}. This dataset comprises 502 open-source ML projects collected from GitHub and analyzed in their study. The selected projects met specific inclusion criteria to ensure both quality and relevance: each project (i) contained at least 100 commits to guarantee a sufficient development history, (ii) had a minimum of two contributors and two stars to ensure community activity and popularity, and (iii) included verifiable ML components such as \texttt{TensorFlow} or \texttt{PyTorch}.
The dataset contains 86{,}143 extracted logging statements, covering diverse ML domains such as computer vision, natural-language processing, and reinforcement learning. In their empirical analysis, the authors identified the use of 12 distinct logging libraries, with 502 projects using at least one logging library, which they categorized into two groups: (i) general-purpose logging libraries (e.g., \texttt{logging}) and (ii) ML-specific logging frameworks (e.g., \texttt{MLflow}, \texttt{TensorBoard}, \texttt{Weights \& Biases}).
In our study, we reused this curated dataset as the foundation for identifying and characterizing logging smells in ML code.

\item \textbf{Checking project on GitHub}: After acquiring the initial list of 502 ML-based GitHub repositories from the dataset provided, we verified that each project remained publicly accessible and active. The primary aim of this step was to curate our corpus by removing projects that are no longer actively maintained. To perform this validation, we developed a Python script leveraging GitHub REST API \footnote{\url{https://docs.github.com/en/rest}}. This automated validation process identified 58 repositories that had been archived by their owners. After filtering out these inactive projects, we proceeded with a final, verified dataset of 444 active repositories for the subsequent stages of our analysis.

\item \textbf{Collecting file containing logging statement}: Following the repository filtering in Step 2, we proceeded by cloning each of the 444 remaining repositories to a local environment. Our goal in this phase was to identify all Python source files that incorporate at least one of the 12 logging libraries. This entailed two sub-tasks: repository cloning and automated static scanning of files.
For the scanning task we developed a Python tool based on the built-in AST \footnote{\url{https://docs.python.org/3/library/ast.html}} (Abstract Syntax Tree) module. The ast library allows programmatic parsing of Python source code into a tree structure that can be traversed to detect import statements and other constructs.  Our script is structured as follows:
\begin{enumerate}
    \item Traverse the directory structure of each cloned repository and locate all files with the \texttt{.py} extension.
    \item For each file, parse the source code using \texttt{ast.parse()}, generating an AST representation of the file’s contents.
    \item Use a custom visitor or walker (e.g., via \texttt{ast.NodeVisitor} or \texttt{ast.walk()}) to inspect nodes of type \texttt{ast.Import} and \texttt{ast.ImportFrom}.
    \item Check whether any of the imported module names or sub-modules correspond to one of the target logging libraries. For instance:
        \begin{itemize}
            \item \texttt{import mlflow} would indicate usage of the \texttt{MLflow} logging framework.
            \item \texttt{import wandb} would indicate usage of the \texttt{Weights\&Biases} framework.
        \end{itemize}
    \item If a file contains at least one import of a targeted logging library, mark it as a \emph{logging-statement file} and retain it for subsequent analysis.
\end{enumerate}
Through this process we extracted 19,775 Python files that each contain at least one import of a logging library from the targeted set.

\item \textbf{Our Json dataset of logging code}: In this step, we converted the filtered set of Python files (from Step 3) into a structured JSON dataset designed to support both quantitative and qualitative analysis. Using the Python ast (Abstract Syntax Tree) module, we parsed each file in order to extract relevant metadata and contextual information about the logging-library usage. Specifically, for each file that imported at least one of the 12 target logging libraries, our tool harvested the following fields:
\begin{itemize}
  \item \texttt{"id"}: a unique identifier for the file. 
  \item \texttt{"project\_name"}: the GitHub repository name from which the file originates.
  \item \texttt{"file\_path"}: the relative path of the source file within the repository. 
  \item \texttt{"library"}: the imported logging library detected in that file (e.g., \texttt{"logging"}).
  \item \texttt{"library\_type"}: classification of the library as either \emph{general-purpose} or \emph{ML-specific} or \emph{Hybrid} i.e, using both logging library type.
  \item \texttt{"python\_file\_content"}: the full text content of the Python source file so that the logging-statements and surrounding context remain available.
  \item \texttt{"snippets"}: a list of code snippets extracted from the file. Each snippet object includes a \texttt{"snippet\_id"}, the exact \texttt{"snippet"} text 
  (for example, \texttt{"logging.warning('Failed to compute shapes: \%s', e)"}) which is the exact logging statement; and contextual metadata such as the \texttt{"class\_name"}, the class where the logging statement is located; \texttt{"function\_name"}, the function where the logging statement is located; \texttt{"line\_number"} where the logging call occurs; \texttt{"line\_before"} the line before the logging statement; and \texttt{"line\_after"} the line after the logging statement.  
\end{itemize}

To provide an overview of our JSON dataset at a glance, Table \ref{tab:dataset-summary} summarizes key statistics of the dataset:

\begin{table}[h]
\centering
\caption{Summary statistics of the JSON corpus of logging-statement files.}
\label{tab:dataset-summary}
\begin{tabular}{lr}
\hline
\textbf{Metric} & \textbf{Count} \\
\hline\hline
Number of Python files & 19,775 \\
Number of general-purpose logging libraries & 12,820 \\
Number of ML-specific logging libraries & 6,431 \\
Number of hybrid usages & 524 \\
Number of functions & 298,182 \\
Number of classes  & 42,000 \\
Number of logging statements & 142,535 \\
\hline
\end{tabular}
\end{table}

In building this dataset we ensured that each file could be traced back to both its repository and logging-library usage, enabling subsequent classification and analysis of logging smells. The corpus generated in this step serves as the primary data source for our RQ1 analysis (taxonomy construction) and supports the qualitative coding of RQ2 (practitioner perceptions).

Furthermore, to support reproducibility and enable ongoing research, we have made the full dataset publicly available via our replication package \citep{replication}.

\end{enumerate} 

\section{Results}
\label{sec:result}
In this section, we present our empirical findings in accordance with the two research questions presented in the introduction. For each question (RQ1 and RQ2), we follow a consistent structure: first we state the motivation of the question, then summarize the analytical approach we applied, and finally report the results we obtained. 

\subsection{\textbf{RQ1: What logging smells exist in open-source ML-based systems?}}
\subsubsection{Motivation:}
Logging is critical for debugging, monitoring, and auditing ML systems, yet improper logging practices can undermine model transparency and system reliability \citep{foalem2025logging}. As in traditional software domains--where recurring bad practices such as comment smells \citep{jabrayilzade2024taxonomy}, linguistic anti-patterns \citep{arnaoudova2013new}, and continuous integration smells \citep{zampetti2020empirical} have been studied--ML logging is also likely to exhibit its own class of code smells that negatively impact maintainability and observability, particularly due to the complexity of AI components. ML systems involve intricate, data-dependent workflows in which logs must capture not only execution states but also model parameters, data lineage, and performance metrics \citep{foalem2024studying}. Inadequate or malformed logs can obscure the root causes of silent failures, hinder reproducibility, and erode the trustworthiness of the system. Addressing this research question is a necessary step toward improving ML observability, maintainability, and system reliability.
\subsubsection{Approach:}
\label{sec:subsectionapprocah}
To answer RQ1, we adopted a mixed-method approach at the function-level and file-level of the codebase. Given the large volume of source files and functions containing logging statements, it is infeasible to manually inspect every element. Our method combines proportional sampling with an iterative, human-in-the-loop analytical process that leverages LLM for initial identification and expert human analysis for validation and refinement. Hence, our approach proceeds in two nested levels of analysis:

\paragraph{Function-level analysis:}  
We begin our investigation at the function level, consistent with methodologies adopted in prior work on logging analysis and automated logging suggestion using artificial intelligence models ~\citep{zhu2015learning, chen2017characterizing, kim2020automatic}. Our goal in this step is to identify logging smells by focusing on how logging statements are structured and distributed within individual functions in ML-based systems.

We started from our JSON dataset, which included 298{,}182 functions extracted from 19{,}775 source files. These files collectively contained 142{,}535 logging statements. At this stage, not all functions necessarily included executable logging statements--some simply housed logger configuration routines or no logging statement.

To isolate logging statements relevant for analysis, we filtered out functions that exclusively contain logging configuration code. These include initialization or setup calls such as \texttt{logging.basicConfig}, \texttt{logging.getLogger}, \texttt{wandb.init}, \texttt{mlflow.set\_tags}, \texttt{neptune.init}, and similar constructs. This filtration step mirrors best practices in prior work~\citep{foalem2025logging}, as configuration statements do not represent actionable observability events at the function level. The complete list of filtered logging configurations is included in our replication package~\citep{replication}.

After applying this filtration, we obtained a refined subset of 15{,}911 logging statements, distributed across 4{,}528 distinct functions. Table~\ref{tab:function-level-stats} summarizes the statistics of logging-library usage and log levels derived from our filtered dataset. This includes the number of functions with at least one logging statement, total logging statements, per-library usage counts, and distribution across standard logging severity levels.
\begin{itemize}
\item Sample selection: We performed stratified sampling at the function level using \citet{cochran1977sampling} formula for proportions to ensure statistical significance and coverage across both general-purpose and ML-specific logging libraries. The following formula was used to determine the sample size for each group:

\[
n_0 = \frac{z_{\alpha/2}^{2} \cdot p (1 - p)}{E^{2}}, \quad n = \frac{n_0}{1 + \dfrac{n_0 - 1}{N}}
\]

where $z_{\alpha/2} = 2.58$ for a 99\% confidence level, $p = 0.5$ is the conservative estimate of the proportion, $E = 0.05$ is the desired margin of error, and $N$ is the population size for each logging group. This method is widely used in empirical software engineering \citep{saarimaki2024taxonomy, zhang2019empirical, wen2021empirical}. Table~\ref{tab:function-sample-99} presents the population size and resulting corrected sample size $n$ for each category. 
\begin{table}[h]
\centering
\begin{threeparttable}
\caption{Sample sizes for function-level logging analysis (99\% confidence, 5\% margin of error).}
\label{tab:function-sample-99}

\begin{tabular}{l r r r r}
\toprule
\textbf{Group} & \textbf{Population $N$} & \textbf{Initial $n_0$} & \textbf{Corrected $n$} & \textbf{Rounded $n$} \\
\midrule
wandb                    & 367  & 665 & $\approx 236.71$ & 237 \\
neptune                  & 137  & 665 & $\approx 113.74$ & 114 \\
tensorflow               & 27   & 665 & $\approx 26.01$  & 26 \\
mlflow                   & 317  & 665 & $\approx 214.93$ & 215 \\
comet\_ml                & 85   & 665 & $\approx 75.49$  & 76 \\
dowel                    & 245  & 665 & $\approx 179.16$ & 179 \\
ml\_logger               & 126  & 665 & $\approx 105.99$ & 106 \\
tensorboard              & 178  & 665 & $\approx 140.59$ & 141 \\
whylogs                  & 3    & 665 & $\approx 2.98$   & 3 \\
sacred                   & 2    & 665 & $\approx 1.99$   & 2 \\
logging (warning+warn)   & 522  & 665 & $\approx 292.69$ & 293 \\
logging (info)           & 1685 & 665 & $\approx 477.03$ & 477 \\
logging (exception)      & 106  & 665 & $\approx 91.50$  & 92 \\
logging (debug)          & 343  & 665 & $\approx 226.70$ & 227 \\
logging (error)          & 357  & 665 & $\approx 231.86$ & 232 \\
logging (fatal)          & 10   & 665 & $\approx 9.94$   & 10 \\
logging (critical)       & 18   & 665 & $\approx 17.54$  & 18 \\
\midrule
\textbf{Total} & \textbf{4,528} & & & \textbf{2,448} \\
\bottomrule
\end{tabular}

\begin{tablenotes}[flushleft]
\footnotesize
\item \textit{Group:} denotes the ML logging library used to categorize the projects.
\item \textit{Population $N$:} represents the total number of functions identified for each group in the dataset.
\item \textit{Initial $n_0$:} is the required sample size computed assuming an infinite population under a 99\% confidence level and a 5\% margin of error.
\item \textit{Corrected $n$:} is the adjusted sample size after applying finite population correction to account for the actual population size $N$.
\item \textit{Rounded $n$:} corresponds to the final sample size used in the study after rounding the corrected value to the nearest integer.
\end{tablenotes}

\end{threeparttable}
\end{table}

\item \emph{Logging smell identification.} We designed an iterative, human-in-the-loop process to identify and build a taxonomy of logging smells. This process was necessary to mitigate the risk of LLM ``hallucinations'' \citep{abbassi2025unveiling, kushwah2025ai}.
We leveraged LLM as an analysis assistant, a technique gaining traction in software engineering for tasks like bug detection and program comprehension \citep{wu2024effective, mamirov2025systematic, lehtinen2024let}. We specifically chose the GPT-5-mini model for its balance of strong reasoning capabilities and cost-efficiency, and this model has a better understanding of logging practice in ML systems than GPT-4o-mini \citep{abbassi2025unveiling, verdet2025assessing}. To ensure reproducible results, the model's temperature was set to 0.2 as recommended \citep{OpenAI, verdet2025assessing}.
We engineered a detailed prompt instructing the LLM to act as an expert in ML observability and software quality. The prompt directed the model to analyze a given Python function, identify at most one logging smell per function if found, then provide a detailed rationale for the identified smell, and assign a confidence level (High, Medium, Low) to its results. Even for a function identified with no smell, the model was instructed to provide similar output. To facilitate further analysis, the LLM was instructed to return its results in a structured JSON format. Our complete prompt is available in our replication package \citep{replication}.

\textbf{Iterative Taxonomy Construction and Validation Process.} Our dataset of 2{,}448 function-level logging instances was divided into three subsets to support a systematic taxonomy construction workflow: 10\% for initial seed derivation, 40\% for iterative refinement and saturation analysis, and the remaining 50\% for evaluating classification reliability. The procedure was carried out by the first two authors--a PhD student with over six years of experience in software engineering and artificial intelligence, and a postdoctoral researcher with more than eight years of expertise in these areas.

\begin{enumerate}[label=(\arabic*), itemsep=0pt, topsep=3pt] 
\item \textbf{Phase 1: Initial Seed Generation (10\%).} We began by selecting a random 10\% of our stratified function-level sample, corresponding to 245 functions out of the total 2,448. These functions were then submitted to the LLM for logging smell identification. The first two authors manually reviewed the LLM outputs, analyzing each function alongside the model's rationale and stated confidence level.

This collaborative review process aimed to establish a reliable initial taxonomy of logging smells. The LLM flagged 201 of the 245 functions as containing at least one logging smell, while 44 were labeled as smell-free. However, many of the LLM-suggested smells were highly specific or semantically overlapping. For example, labels such as \texttt{AMBIGUOUS\_LOG\_MESSAGES}, \texttt{AMBIGUOUS\_METRIC\_LOGGING}, and \texttt{AMBIGUOUS\_UNSTRUCTURED\_LOG\_MESSAGES} were deemed to reflect variations of a broader concept, which we ultimately categorized under a more general label, \texttt{AMBIGUOUS}.

Using an open coding strategy \citep{khandkar2009open}, the authors iteratively grouped and refined the 201 LLM-proposed smell labels. This process yielded a seed taxonomy of 11 distinct logging smell types. Of the 201 LLM-suggested cases, 91 were accepted as smells and assigned one of the 10 labels, while 110 were rejected as not being smells because they lacked sufficient context (e.g., some instances were flagged as smells despite lacking adequate information). Among the 44 functions marked by the LLM as having no smell, only one was reclassified by the authors as actually containing a logging smell; the remaining 43 were validated as clean.
This seed generation step served as the foundation for the subsequent iterative refinement and saturation phases.
\item \textbf{Phase 2: Iterative Refinement and Saturation (40\%).} From the remaining sample (2,203 instances), we selected an additional batch of 40\% (441 instances) for further analysis. The half of 40\% is used for LLM-assisted taxonomy expansion, and the other half is used to evaluate saturation. 

\begin{itemize}
          \item \textbf{2a -- LLM-Assisted Taxonomy Expansion (20\%):} We analyzed this subsample using an enhanced prompt that incorporated concise summaries of the smell categories from our seed taxonomy, given the token limitations of the LLM. The model was instructed either to classify each instance into one of the existing categories or to propose a new category when none of the existing ones applied. The goal of this phase was to assess whether additional smell categories would emerge or whether theoretical saturation had been reached.
          The LLM initially proposed 16 new labels for 19 code snippets. The first two authors jointly reviewed all suggestions during a calibration meeting, examining both the corresponding code snippets and the LLM-generated rationales. Through this review, we determined that 11 of the proposed categories did not constitute actual smells, and several others required an execution context not available through static analysis alone. Of the remaining proposals, 6 were ultimately subsumed under 3 previously identified categories--\textit{Log Without Context}, \textit{Ambiguous Logging}, and \textit{Heavy Data Logging}. Finally, 2 proposals were judged to represent a new smell type, which we labeled \textit{Misleading Logging}, and we subsequently developed a formal definition for this new category and also refined the description of \textit{Ambiguous Logging}. Following this refinement step, we reviewed the remainder of the sample to verify whether the LLM assigned the correct category among the 11 established categories, based on its rationale and the provided code snippet. 
          \item \textbf{2b -- LLM-Assisted Taxonomy Saturation Evaluation (20\%):} In this step, we analyzed an additional 20\% (352) of the remaining 1,759 samples using an enhanced prompt that incorporated the newly identified categories from the previous phase. We applied the same prompting strategy as before, instructing the LLM to classify each code snippet into one of the 12 existing categories or to propose a new category if none applied. For five snippets, the LLM suggested three new labels; however, after reviewing these cases through a calibration meeting, all were ultimately classified as \emph{no smell}. Since no additional distinct categories emerged at this stage, we concluded that the taxonomy had reached saturation. We then proceeded to manually review the rest of the sample to verify whether the LLM assigned the appropriate category among the 12 established categories, based on its rationale and the provided code snippet. 
\end{itemize}
\item \textbf{Phase 3: Reliability Assessment and Final Labeling (50\%).} At this stage, we instructed the LLM to classify the remaining 1{,}410 samples into one of the 12 identified categories, without allowing new category suggestions, as saturation had been reached in the previous phase. The first two authors independently reviewed all LLM-generated labels to evaluate both their own understanding of the categories and the reliability of the overall labeling process. We then computed inter-rater agreement based on the two coders’ independent modifications of the LLM-produced labels, obtaining a $\kappa = 0.816755$, indicating an almost perfect level of agreement. All remaining disagreements were subsequently resolved through discussion. 
\end{enumerate}
\end{itemize}

\begin{table}[h]
\centering
\caption{Statistics of the Filtered Dataset for RQ1 Analysis}
\begin{adjustbox}{width=0.35\textwidth}
\label{tab:function-level-stats}
\begin{tabular}{l r}
\toprule
\textbf{Statistic} & \textbf{Count} \\
\midrule
\textbf{Total Unique Functions} & \textbf{4,528} \\
\textbf{Total Logging Statements} & \textbf{15,911} \\
\midrule
\multicolumn{2}{l}{\textit{Logging Statements by Library}} \\
logging     & 10,499 \\
dowel       & 2,031 \\
wandb       & 1,459 \\
mlflow      & 811 \\
ml\_logger  & 369 \\
tensorboard & 320 \\
neptune     & 240 \\
comet\_ml   & 108 \\
tensorflow  & 69 \\
whylogs     & 3 \\
sacred      & 2 \\
\midrule
\multicolumn{2}{l}{\textit{Logging Statements by General Log Level}} \\
info        & 6,625 \\
warning (includes `warn`) & 1,471 \\
debug       & 1,315 \\
error       & 895 \\
exception   & 150 \\
critical    & 27 \\
fatal       & 16 \\
\midrule
\multicolumn{2}{l}{\textit{Unique Functions by Library or Log Level}} \\
wandb         & 367 \\
neptune       & 137 \\
tensorflow    & 27 \\
mlflow        & 317 \\
comet\_ml     & 85 \\
dowel         & 245 \\
ml\_logger    & 126 \\
tensorboard   & 178 \\
whylogs       & 3 \\
sacred        & 2 \\
logging (warning)   & 445 \\
logging (info)      & 1,685 \\
logging (exception) & 106 \\
logging (debug)     & 343 \\
logging (error)     & 357 \\
logging (warn)      & 77 \\
logging (fatal)     & 10 \\
logging (critical)  & 18 \\
\bottomrule
\end{tabular}
\end{adjustbox}
\begin{tablenotes}
\small
\item \textit{Note.} Log levels are only for the general-purpose logging library (e.g., \texttt{logging}, \texttt{warnings}). Unique function counts reflect functions with at least one logging statement, by library or log level.
\end{tablenotes}
\end{table}

\subsubsection{\textbf{Result:}}
\begin{figure}
    \centering
    \includegraphics[width=0.9\linewidth]{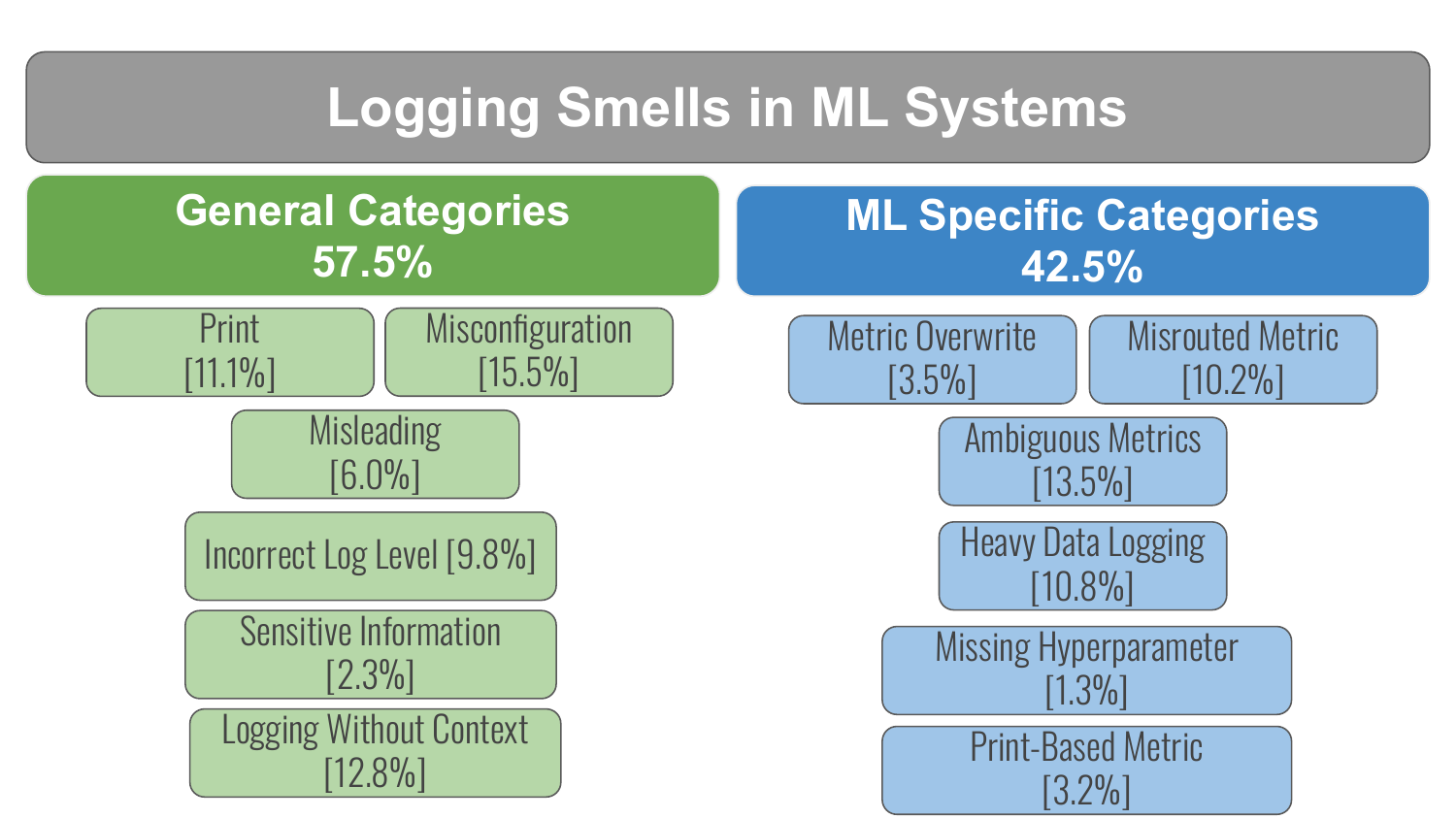}
    \caption{Taxonomy of Logging Smells in ML Systems.}
    \label{fig:placeholder}
\end{figure}

This section presents the 12 logging-related smell categories identified in our study. These smells capture recurring misuse and suboptimal practices in both general-purpose and ML-specific logging within machine learning codebases.\\
Figure \ref{fig:placeholder} presents the taxonomy of logging smells identified in our study, distinguishing between general categories and ML-specific categories. The general categories (71\%) correspond to logging smells that are well-known in the software engineering literature and that arise in non-ML components of ML systems, reflecting issues common to traditional software (e.g., ambiguous logs, misconfiguration, etc). In contrast, the ML-specific categories (29\%) capture smells that originate from ML components themselves—such as metric misuse, missing hyperparameters, or improper experiment-tracking practices—highlighting logging challenges unique to data-driven and model-centric workflows. This taxonomy illustrates both the shared and domain-specific nature of logging problems within ML-based software systems.\\

\textbf{Ambiguous Metrics Logging} logging arises when logged metrics or values lack sufficiently clear semantic descriptions, making it difficult to understand what is being measured or how the results should be interpreted during auditing. Ambiguity also occurs when the descriptive label of a metric does not correspond to the actual value being logged (e.g., \texttt{mlflow.log\_metric("train\_{}".format(metric), valid\_loss)}), where the recorded value contradicts the metric’s intended meaning. In this example, the log entry suggests that the value reflects training performance (because the metric name is prefixed with “train”), yet the value actually logged corresponds to the validation loss. This mismatch introduces a semantic contradiction between the metric’s name and its content, potentially misleading both practitioners and automated analysis tools. More generally, ambiguity arises when metric names or log messages fail to provide enough contextual information to clarify their meaning, scope, or statistical role. The problem is further exacerbated when metrics are logged without any descriptive message at all. As illustrated in Listing~\ref{lst:Ambiguous}, the code logs \texttt{x} without offering any contextual explanation to indicate what the value represents or how it should be interpreted.\\
\begin{lstlisting}[style=python, caption={Example of an Ambiguous Logging Smell.}, label={lst:Ambiguous}]
        def create_and_log_run(name):
            run_id = self.store.create_run(
                experiment_id,
                user_id="MrDuck",
                start_time=123,
                tags=[entities.RunTag(mlflow_tags.MLFLOW_RUN_NAME, name)]).info.run_id
            self.store.log_metric(run_id, entities.Metric("x", float(name), 1, 0))
            return run_id
\end{lstlisting}

\textbf{Misleading Logging} refers to log messages that inaccurately describe the system's state, the operations performed, or the conditions under which those operations occur. Unlike ambiguous logs, which suffer from insufficient context or vague terminology, misleading logs are problematic because they \emph{communicate information that is factually incorrect, incomplete, or not guaranteed to be true}. This form of logging produces observability signals that conflict with actual program behaviour, thereby eroding the reliability of logs as an auditing or debugging tool.
As illustrated in Listing~\ref{lst:Misleading}, which is a \texttt{load\_pretrained snippet}, a logging statement is emitted at line 17 unconditionally, even though the resizing operation occurs only if the subsequent \texttt{if} condition is satisfied. The contrast can be observed with the next \texttt{if} condition. 
Thus, the log claims an operation has taken place regardless of whether it actually does--making the log misleading. While prior work \citep{saarimaki2024taxonomy} does not capture cases where log statements are inconsistent with the program's control flow. We therefore introduce Misleading Logging to characterize logs that assert operations or states that are not guaranteed to occur during execution.\\
\begin{lstlisting}[style=python, caption={Example of Misleading Logging Smell.}, label={lst:Misleading}]
def load_pretrained(
    model,
    cfg=None,
    num_classes=1000,
    in_chans=3,
    filter_fn=None,
    img_size=224,
    num_frames=8,
    num_patches=196,
    attention_type="divided_space_time",
    pretrained_model="",
    strict=True,
):


    ## Resizing the positional embeddings in case they don't match
    logging.info(
        f"Resizing spatial position embedding from {state_dict['pos_embed'].size(1)} to {num_patches + 1}"
    )
    if num_patches + 1 != state_dict["pos_embed"].size(1):
        pos_embed = state_dict["pos_embed"]
        cls_pos_embed = pos_embed[0, 0, :].unsqueeze(0).unsqueeze(1)
        other_pos_embed = pos_embed[0, 1:, :].unsqueeze(0).transpose(1, 2)
        new_pos_embed = F.interpolate(
            other_pos_embed, size=(num_patches), mode="nearest"
        )
        new_pos_embed = new_pos_embed.transpose(1, 2)
        new_pos_embed = torch.cat((cls_pos_embed, new_pos_embed), 1)
        state_dict["pos_embed"] = new_pos_embed

    ## Resizing time embeddings in case they don't match
    if "time_embed" in state_dict and num_frames != state_dict["time_embed"].size(1):
        logging.info(
            f"Resizing temporal position embedding from {state_dict['time_embed'].size(1)} to {num_frames}"
        )
        time_embed = state_dict["time_embed"].transpose(1, 2)
        new_time_embed = F.interpolate(time_embed, size=(num_frames), mode="nearest")
        state_dict["time_embed"] = new_time_embed.transpose(1, 2)
\end{lstlisting}



\textbf{Heavy Data Logging} refers to the inclusion of computationally expensive operations within logging routines, such as performing additional model inferences, executing matrix operations, or processing large batches of data. Embedding such heavy computations inside log statements slows down the logging process itself and can significantly impact the overall runtime of training or evaluation, and may even cause bugs~\citep{MLFLOWBug}. As illustrated in Listing~\ref{lst:heavy}, the logging code computes multiple aggregate statistics (e.g., mean, max, and abs) over epochs within the logging function itself. When such logging routines are executed frequently (e.g., per iteration or per epoch), these repeated matrix and vector operations can introduce non-negligible overhead. \\
\begin{lstlisting}[style=python, caption={Example of Heavy Data Logging Smell.}, label={lst:heavy}]
 def _log_statistics(self):
        """Output training statistics to dowel such as losses and returns."""
        tabular.record('Policy/AveragePolicyLoss',
                       np.mean(self._episode_policy_losses))
        tabular.record('QFunction/AverageQFunctionLoss',
                       np.mean(self._episode_qf_losses))
        tabular.record('QFunction/AverageQ', np.mean(self._epoch_qs))
        tabular.record('QFunction/MaxQ', np.max(self._epoch_qs))
        tabular.record('QFunction/AverageAbsQ',
                       np.mean(np.abs(self._epoch_qs)))
        tabular.record('QFunction/AverageY', np.mean(self._epoch_ys))
        tabular.record('QFunction/MaxY', np.max(self._epoch_ys))
        tabular.record('QFunction/AverageAbsY',
                       np.mean(np.abs(self._epoch_ys)))
\end{lstlisting}

\textbf{Misconfigured Logging} captures improper or inconsistent initialization of logging frameworks, particularly when logging configuration is embedded directly within core program logic rather than being centralized or externally defined. Centralizing logging configuration is widely regarded as a best practice, as it promotes consistency, reuse, and maintainability~\citep{zhi2019exploratory, BetterStack}. Because logging configuration is global and stateful, configuring it conditionally within the execution path can lead to unpredictable behavior, conflicting log-level settings, or silent overrides of the intended configuration, especially when the code is reused, imported, or executed in different contexts.
Listing~\ref{lst:misconfigured} illustrates this smell. In this example, logging configuration is performed inside the \texttt{main} function and varies based on runtime arguments. Such in-code configuration tightly couples logging behavior to application logic, making it difficult to reason about which configuration is ultimately applied and increasing the risk of inconsistent logging behavior across runs.\\
\begin{lstlisting}[style=python, caption={Example of Misconfigured Logging Smell.}, label={lst:misconfigured}]
def main(args):
    ...
    if args.verbose == 1:
        logging.basicConfig(
            level=logging.INFO,
            format="%(asctime)s (%(module)s:%(lineno)d) %(levelname)s: %(message)s",
        )
    elif args.verbose == 2:
        logging.basicConfig(
            level=logging.DEBUG,
            format="%(asctime)s (%(module)s:%(lineno)d) %(levelname)s: %(message)s",
        )
    else:
        logging.basicConfig(
            level=logging.WARN,
            format="%(asctime)s (%(module)s:%(lineno)d) %(levelname)s: %(message)s",
        )
        logging.warning("Skip DEBUG/INFO messages")

    # check input arguments
    if not os.path.exists(args.raw_trans) and args.stage > 0:
        logging.error("Filename '" + args.raw_trans + "' is not exist")
        logging.error("Please check the '--raw-trans' argument")
        sys.exit(1)
    else:
        logging.info("Loading raw text: " + args.raw_trans)

    # make logging dir
    if not os.path.exists(args.log_dir):
        os.makedirs(args.log_dir)

    # run refining processes
    logging.debug(
        "stage 1: Remove meta-symbols for unwanted pause and elongated segment"
    )
    in_fn = args.raw_trans
    out_fn = "1_remove_meta_symbols.txt"
    if args.raw_trans is not None and args.stage <= 1:
        logging.debug(">> reading %s", in_fn)
        with codecs.open(in_fn, mode="r", encoding="utf-8") as f:
            trans_raw = f.readlines()
    ...
\end{lstlisting}


\textbf{Misrouted Metric Logging} This smell denotes cases where ML performance metrics or runtime indicators are logged through the wrong channel-typically using a general-purpose logger (\texttt{logging.info}) instead of an ML experiment tracker (e.g., MLflow, W\&B, Neptune). Misrouted metrics are not captured in experiment dashboards, cannot be compared across runs, and undermine traceability and governance requirements for ML system development. Listing~\ref{lst:Misrouted} illustrates this smell. In the example, the test loss is correctly logged to W\&B using \texttt{wandb.log}, ensuring it is tracked as part of the experiment. In contrast, the test accuracy--an equally critical evaluation metric--is emitted only via \texttt{logger.info} as part of a formatted log message. As a result, accuracy is not recorded by the ML tracking system and is lost for downstream analysis, visualization, or comparison across runs. This inconsistent routing of metrics across logging channels leads to incomplete and fragmented experiment records.\\
\begin{lstlisting}[style=python, caption=Example of Misrouted Metric Logging Smell., label={lst:Misrouted}]
def test(model, test_loader, device):
    model.eval()
    test_loss = 0
    correct = 0
    with torch.no_grad():
        for data, target in test_loader:
            data, target = data.to(device), target.to(device)
            output = model(data)
            test_loss += F.nll_loss(output, target, size_average=False).item()  # sum up batch loss
            pred = output.max(1, keepdim=True)[1]  # get the index of the max log-probability
            correct += pred.eq(target.view_as(pred)).sum().item()

    test_loss /= len(test_loader.dataset)
    wandb.log({"testing/loss": test_loss})
    logger.info(
        "Test set: Average loss: {:.4f}, Accuracy: {}/{} ({:.0f}%)\n".format(
            test_loss, correct, len(test_loader.dataset), 100.0 * correct / len(test_loader.dataset)
        )
    )
\end{lstlisting}

\textbf{Metric Overwrite} sometimes manifested as duplicate metric logging, occurs when the same metric key is logged multiple times without an explicit step or timestamp, thereby relying on implicit ``last value wins'' semantics. This practice discards valuable historical information, obscures temporal trends, and produces incomplete or misleading experiment traces. As a result, it reduces the ability to diagnose training dynamics, compare checkpoints, or audit model performance over time.
Listing~\ref{lst:metric-overwrite} illustrates this smell. In the example, the metric key \texttt{name\_1} is logged twice within the same MLflow run, first with a value of \texttt{25} and later with \texttt{30}, without specifying a step or timestamp. Consequently, the earlier value is silently overwritten, and the experiment retains only the final value. \\
\begin{lstlisting}[style=python, caption={Example of Metric Overwrite Logging Smell.}, label={lst:metric-overwrite}]
def log_metric():
    try:
        tracking.set_tracking_uri(tempfile.mkdtemp())
        active_run = start_run()
        run_uuid = active_run.info.run_uuid
        with active_run:
            mlflow.log_metric("name_1", 25)
            mlflow.log_metric("name_2", -3)
            mlflow.log_metric("name_1", 30)
            mlflow.log_metric("nested/nested/name", 40)
    ...
\end{lstlisting}


\textbf{Log Without Context} This smell arises when variables or metric values are logged without descriptive messages or labels, leaving log consumers unable to infer their meaning or provenance. Contextless logs negatively affect interpretability, hinder debugging efforts, and complicate automated log parsing pipelines that rely on structured or semantically rich messages. Listing~\ref{lst:log-without-context} illustrates this smell. In the example, exceptions are logged using generic messages such as \texttt{"fail"} or by directly emitting the string representation of the exception via \texttt{logging.error(msg)}. These log statements provide no information about which distribution or version failed, what operation was attempted, or under which conditions the error occurred. As a result, the logs offer limited diagnostic value and make post-mortem analysis significantly more difficult, particularly in concurrent or large-scale execution settings. While prior work \citep{saarimaki2024taxonomy} captures vague or missing log messages under ``Message Madness'' and ``missing identifiers'' under ``Undercover Identifier'', these categories do not explicitly emphasize the absence of contextual information required for interpretation. We therefore refine this dimension through Log Without Context, which highlights the importance of semantic completeness in log messages.\\
\begin{lstlisting}[style=python, caption=Example of Log Without Context Logging Smell., label={lst:log-without-context}]
def _get(distribution, version):
            ...
            for download in [True] if self.download else [False, True]:
                failure = None
                try:
                    result = get_wheel(
                        ...
                    )
                    if result is not None:
                        break
                except Exception as exception:  # noqa
                    logging.exception("fail")
                    failure = exception
            if failure:
                if isinstance(failure, CalledProcessError):
                    ...
                else:
                    msg = repr(failure)
                logging.error(msg)
                with lock:
                    fail[distribution] = version
            else:
                with lock:
                    name_to_whl[distribution] = result
\end{lstlisting}

\textbf{Missing Hyperparameter Logging} refers to the failure to record all hyperparameters that govern an experiment, particularly when some parameters are conditionally logged, omitted by default, or embedded implicitly within the training code. Because hyperparameters directly influence model behavior and learning dynamics, incomplete logging compromises reproducibility, reduces comparability across runs, and weakens auditability within ML governance pipelines. Listing~\ref{lst:missing-hparam} illustrates this smell. In the example, several hyperparameters (e.g., number of epochs, number of classes, optimizer and scheduler configurations) are passed through the \texttt{hyper\_params} dictionary and the training routine but are never explicitly logged to the experiment tracker. Instead, only performance metrics such as loss and accuracy-related measures are recorded. As a result, the experiment metadata captured by the tracking system is incomplete, making it difficult to reproduce the training setup or meaningfully compare results across different runs. 
\\
\begin{lstlisting}[style=python, caption={Example of Missing Hyperparameter Logging Smell.}, label={lst:missing-hparam}]
def train_stagewise(hyper_params, teacher, student, sf_teacher, sf_student, trainloader, valloader, args):
    ...
    for epoch in range(hyper_params['num_epochs']):
        student, highest_iou, train_loss, val_loss, avg_iou, avg_pixel_acc, avg_dice_coeff = train(
            model=student,
            train_loader=trainloader,
            val_loader=valloader,
            num_classes=hyper_params['num_classes'],
            loss_function=criterion,
            optimiser=optimizer,
            scheduler=scheduler,
            epoch=epoch,
            num_epochs=hyper_params['num_epochs'],
            savename=savename,
            highest_iou=highest_iou,
            args=args
        )
        if args.api_key:
            experiment.log_metric('train_loss', train_loss)
            experiment.log_metric('val_loss', val_loss)
            experiment.log_metric('avg_iou', avg_iou)
            experiment.log_metric('avg_pixel_acc', avg_pixel_acc)
            experiment.log_metric('avg_dice_coeff', avg_dice_coeff)
\end{lstlisting}


\textbf{Print-Based Metrics} Print-based metrics capture situations where model performance indicators (e.g., accuracy, loss, reward curves) are emitted using \texttt{print()} rather than an ML tracking library. Print statements produce unstructured, ephemeral outputs that are not traceable, not persisted in experiment repositories, and easily lost during distributed or cloud-based execution. This smell reflects a mismatch between the importance of the metric and the inadequacy of the logging mechanism used to report it. Listing~\ref{lst:PrintMetrics} illustrates this smell. In the example, evaluation metrics (\texttt{test\_dimscore\_*}) computed over the test set are printed directly to standard output using \texttt{print}, rather than being logged to the configured experiment tracker (e.g., Neptune). As a result, critical evaluation information is effectively discarded, undermining experiment traceability and reproducibility.\\
\begin{lstlisting}[style=python, caption=Example of Print-Based Metrics Logging Smell., label={lst:PrintMetrics}]
def saoto_bsearch(seed, neural_net, scoring, expt_name: str, search_space: dict, x_train, y_train, wdir="./",
                  use_neptune=True, x_test=None, y_test=None, tags: [str] = []):
    set_seed(seed)
    neural_net.verbose = 0
    ...
    # gpu mem < 8GB
    if use_neptune:
    ...    
    if x_test is not None and y_test is not None:
        y_pred = opt.predict(x_test)
        for mode in DimScoreModes.keys():
            dims = DimScore(y_test, y_pred, mode)
            print("test_dimscore_{}".format(mode), dims)

    dump(opt, "{}.pkl".format(expt_name))
    if use_neptune:
        sk_utils.log_results(opt.optimizer_results_[-1], experiment=experiment)
        neptune.stop()
    os.chdir(whereami)
\end{lstlisting}

\textbf{Print Logging} Print logging describes the broader misuse of \texttt{print()} as a substitute for a general-purpose logging framework. Unlike structured logs, print statements lack log levels, timestamps, formatting control, and configurability, making them unsuitable for production ML systems. Their presence often signals immature observability design and complicates integration with monitoring, debugging, and governance tools. Listing~\ref{lst:PrintLogging} illustrates this smell. In the example, an error related to the configuration of \texttt{comet\_ml} (i.e., an ML-specific logging library) is reported using a raw \texttt{print()} statement. Because this message bypasses the logging framework, it is emitted without severity information (e.g., \texttt{ERROR}, \texttt{EXCEPTION}), cannot be captured by log collectors, and may be missed entirely in distributed or automated execution environments. Using structured logging instead would allow the message to be consistently recorded, filtered, and audited alongside other system events. While prior work \citep{saarimaki2024taxonomy} captures issues related to missing logs under ``Logging Lost in the Wind'' and ``formatting issues'' under ``Format Turmoil'', these categories do not explicitly account for cases where developers bypass logging frameworks entirely. We therefore introduce Print Logging to capture the use of standard output mechanisms (e.g., \texttt{print()}) as a substitute for structured logging.
\\
\begin{lstlisting}[style=python, caption=Example of Print Logging Smell., label={lst:PrintLogging}]
    def __init__(self):
        ...
        if comet_installed:
            try:
                ...
            except:
                print("ERROR: comet_ml is installed, but not configured properly (e.g. check API key setup). HTML reports will not be uploaded.")
\end{lstlisting}

\textbf{Logging Sensitive Data} This smell denotes instances in which sensitive, personally identifiable, or otherwise confidential information is inadvertently recorded in logs, including security-critical elements such as API keys, authentication tokens, or access credentials. Logging such sensitive data introduces substantial security, privacy, and compliance risks, particularly in regulated domains or applications subject to frameworks such as the GDPR \citep{voigt2017eu} or the AI Act \citep{AIActEU2023}. Listing~\ref{lst:Sensitive} illustrates this smell. In the example, the configuration dictionary \texttt{config} contains a \texttt{comet\_api\_key}, which is subsequently passed to the experiment tracker and logged wholesale via \texttt{log\_multiple\_params}. As a result, the API key may be persisted in experiment metadata, dashboards, or backend storage, unintentionally exposing a security credential. Such practices underscore the need for principled logging strategies that explicitly filter, redact, or exclude protected information before it is recorded. Prior work \citep{saarimaki2024taxonomy} on logging smells has primarily focused on message quality, and structure, while treating security risks such as information leakage as consequences rather than first-class smells. We therefore introduce Logging Sensitive Data as a distinct category, capturing cases where confidential or personally identifiable information is recorded in logs, leading to security and compliance violations.
\\
\begin{lstlisting}[style=python, caption=Example of Sensitive Logging Smell., label={lst:Sensitive}]
    def __init__(self, sess, summary_dir, config, scalar_tags=None, images_tags=None):
        self.init_summary_ops()

        self.summary_writer = tf.summary.FileWriter(summary_dir)

        if "comet_api_key" in config:
            from comet_ml import Experiment
            self.experiment = Experiment(api_key=config['comet_api_key'], project_name=config['exp_name'])
            self.experiment.log_multiple_params(config)
\end{lstlisting}

\textbf{Incorrect Log Level} 
Incorrect log level refers to situations in which events are logged with a severity level that does not accurately reflect their importance or impact on program execution. This smell commonly occurs when critical failures, exceptions, or abnormal behaviors are logged using low-severity levels (e.g., \texttt{INFO}) instead of more appropriate levels such as \texttt{WARNING} or \texttt{ERROR}. As a result, important runtime issues may be overlooked, filtered out, or misinterpreted during debugging and monitoring.

Listing~\ref{lst:incorrect-level} illustrates this smell in the context of ML experiment logging. When an exception occurs during a call to \texttt{wandb.log}, the error is caught and reported using \texttt{LOGGER.info}, even though the failure affects the experiment’s logging backend and may compromise result traceability. Moreover, because the exception is logged at an informational level and without a stack trace, the severity of the issue is understated and its root cause is difficult to diagnose. Such misuse of log levels reduces observability, weakens failure detection mechanisms, and can lead to silent degradation of experiment tracking and auditing capabilities. This smell directly corresponds to the ``Mercurial Logging Level'' category identified in prior work \citep{saarimaki2024taxonomy}, which captures the misuse of logging severity levels. In our context, we further highlight its impact on observability and failure detection, particularly in ML pipelines where incorrect log levels can obscure critical issues affecting experiment traceability.
\begin{lstlisting}[style=python, caption=Example of Incorrect Log Level Smell., label={lst:incorrect-level}]
    def end_epoch(self, best_result=False):
        """
        commit the log_dict, model artifacts and Tables to W&B and flush the log_dict.

        arguments:
        best_result (boolean): Boolean representing if the result of this evaluation is best or not
        """
        if self.wandb_run:
            with all_logging_disabled():
                if self.bbox_media_panel_images:
                    self.log_dict["BoundingBoxDebugger"] = self.bbox_media_panel_images
                try:
                    wandb.log(self.log_dict)
                except BaseException as e:
                    LOGGER.info(
                        f"An error occurred in wandb logger. The training will proceed without interruption. More info\n{e}"
                    )
	...
\end{lstlisting}

\begin{tcolorbox}[colback=black!4,colframe=black!50!white]
\textbf{Findings:}
We identified 12 categories of logging smells in ML code. These categories span both \textbf{general-purpose logging smells}—including \textit{Misconfiguration, Misleading, Print, Incorrect Log Level, Sensitive Data,} and \textit{Log Without Context}—which align with or extend existing logging smell taxonomies, and a set of \textbf{ML-specific logging smells} that are not captured in prior work. 
In particular, we identify 6 \textbf{ML-specific categories}—\textit{Ambiguous Metrics Logging, Metric Overwrite, Misrouted Metric Logging, Heavy Data Logging, Missing Hyperparameter Logging,} and \textit{Print-Based Metrics}—which arise from the unique characteristics of ML pipelines, such as experiment tracking, metric management, and data-intensive workflows. 
\end{tcolorbox}

\subsection{\textbf{RQ2: How do ML practitioners perceive and experience these logging smells in practice?}}
\subsubsection{Motivation:}
While identifying logging smells is essential for improving ML code quality, it is equally important to understand how practitioners perceive and experience these smells. Developers' awareness, interpretation, and tolerance of such smells can vary significantly depending on their domain expertise, project constraints, and deployment environments. Prior work in software engineering has shown that developers’ perceptions influence how smells are prioritized, mitigated, or even ignored in practice \citep{zampetti2020empirical,arnaoudova2013new}. By investigating practitioners' experiences with the ML logging smells identified in RQ1, this research question aims to externally validate the relevance and impact of these smells in practical development contexts.\\
\subsubsection{Approach:}
To investigate how ML practitioners perceive and experience the logging smells identified in Phase~1, we conducted a structured online survey following established methodologies for evaluating software engineering taxonomies \citep{jabrayilzade2024taxonomy, abbassi2025unveiling}. Our objective was to assess the perceived relevance, frequency, and severity of each smell type and to gather qualitative insights that contextualize practitioners’ real-world experiences with logging in ML workflows.
    
    \textit{Survey Design.} The survey was administered via Google Forms\footnote{\url{https://docs.google.com/forms/u/0/}} and consisted of 4 sections. 
    First, a welcome page introduced the study objectives, outlined the potential benefits of improving ML logging practices, and obtained informed consent. 
    Second, a demographic section collected respondents’ professional backgrounds, including their job roles and years of experience in software engineering, Python programming, and machine learning. We also asked about the frequency with which they add or modify logging statements and the logging libraries they use. 
    Finally, the core validation section presented the catalog of logging smells derived in Phase~1. Each smell was introduced with (i) a concise textual definition and (ii) a representative code snippet extracted from our dataset. For each smell, participants provided ratings using a 6-point Likert scale for \emph{relevance} (importance of addressing the smell), \emph{frequency} (how often they encounter it), and \emph{severity} (its impact on maintainability, observability, and reproducibility). In addition, participants indicated their level of agreement with the proposed smell classification using a 3-point Likert scale (Neutral, Disagree, Agree).
    
    The survey included the following questions for each smell:
    \begin{enumerate}
        \item Do you agree that this constitutes a logging smell? [Likert scale]
        \item How frequently have you observed this smell in your projects? [Likert scale]
        \item How relevant is it to address this smell in practice? [Likert scale]
        \item If this smell occurs, how severe are its negative consequences? [Likert scale]
        \item In your own words, what makes this logging smell problematic in practice? Please describe any specific technical issues, failures, or challenges you have personally encountered when dealing with this smell. [Open-ended]
    \end{enumerate}
    
    At the end of the survey, participants could optionally provide additional comments and report logging smells not included in our catalog.
    
    \textit{Participant Recruitment.} Consistent with prior software engineering survey research \citep{jabrayilzade2024taxonomy, abbassi2025unveiling, foalem2024studying, foalem2025logging}, we targeted practitioners directly involved in ML development. Recruitment relied on contributors to the 444 open-source ML projects analyzed in Phase~1, whose publicly available email addresses were collected using the GitHub API. As an incentive, participants were offered entry into a raffle for a \$100 Amazon gift card.\\
    
    \textit{Pilot Testing.} Before launching the full survey, we conducted a pilot study with 5 ML engineers contacted by the first two authors to assess clarity, completion time, and the interpretability of the smell descriptions. Based on their feedback, we refined the wording of several questions to improve clarity and politeness, enhanced examples for potentially ambiguous smells, improved the overall descriptions of logging issues, added a ``less than one year'' option in the experience section, and adjusted the estimated completion time to approximately 30 minutes. Responses from the pilot study were excluded from the final analysis.\\ 
\subsubsection{Results:}
\begin{figure}[t]
        \centering
        \includegraphics[width=\textwidth]{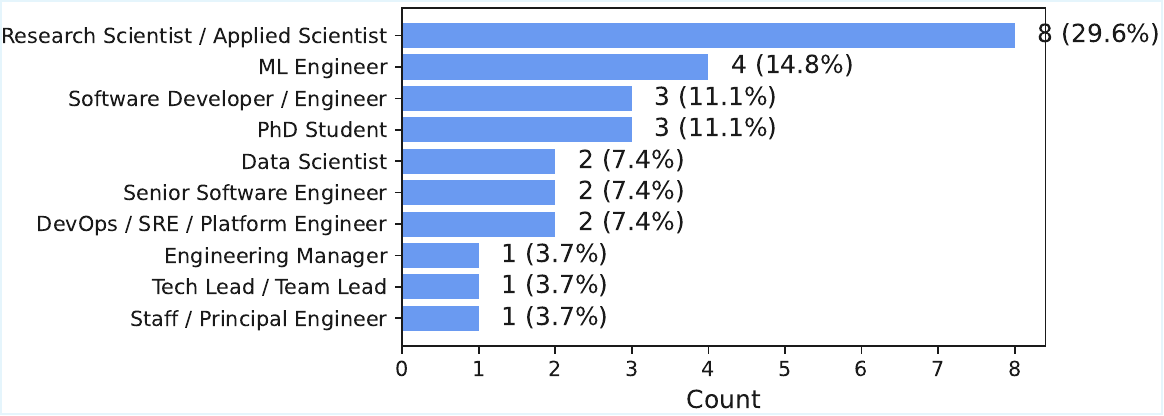}
        \caption{Professional roles of survey respondents.}
        \label{fig:participant_roles}
    \end{figure}

\begin{figure*}[t]
        \centering
        
        \begin{subfigure}[b]{0.4\textwidth}
            \centering
            \includegraphics[width=\textwidth]{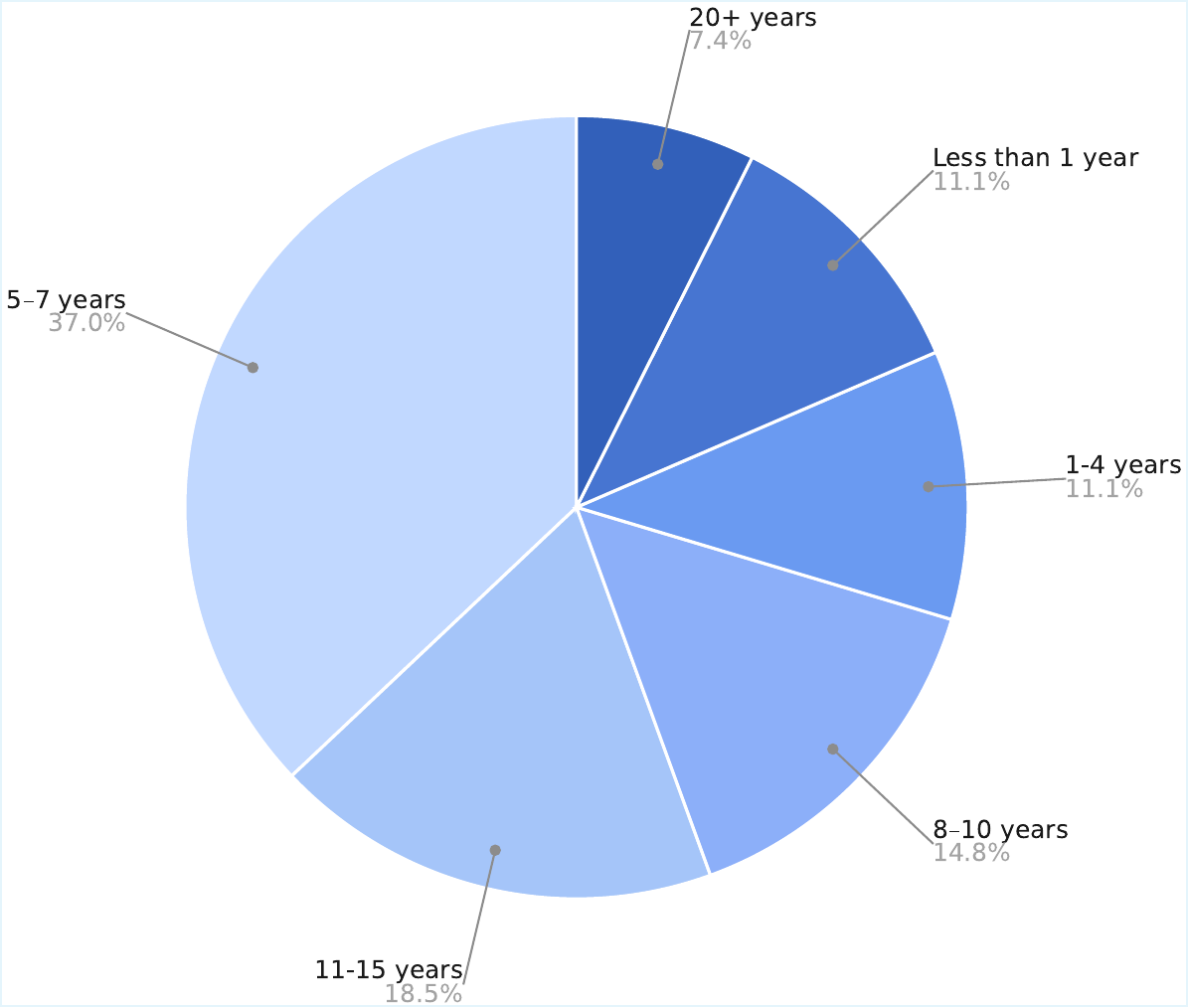}
            \caption{Software Engineering}
            \label{fig:se_experience}
        \end{subfigure}
        \hfill
        \begin{subfigure}[b]{0.4\textwidth}
            \centering
            \includegraphics[width=\textwidth]{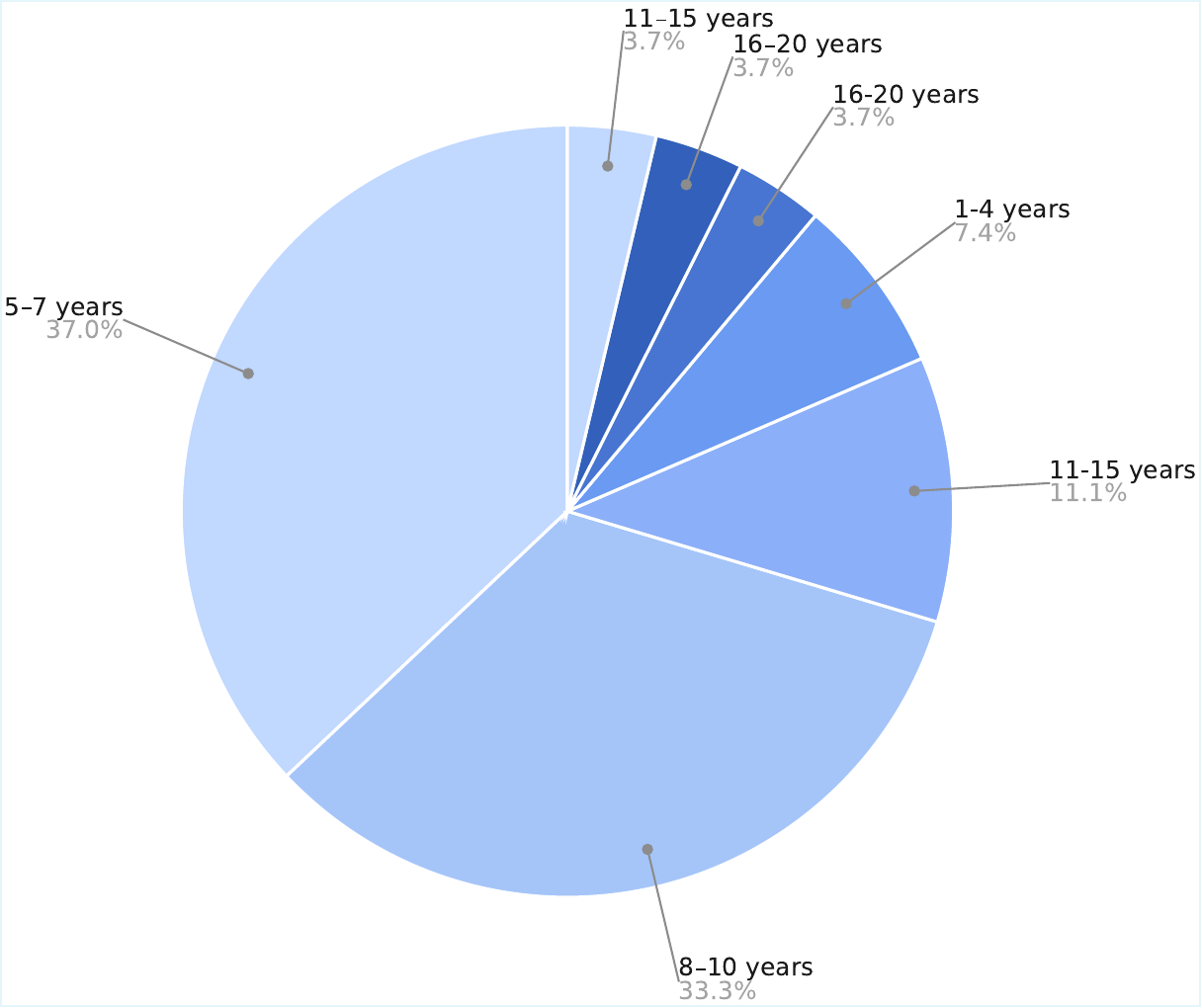}
            \caption{Python Programming}
            \label{fig:python_experience}
        \end{subfigure}
        \hfill
        \begin{subfigure}[b]{0.4\textwidth}
            \centering
            \includegraphics[width=\textwidth]{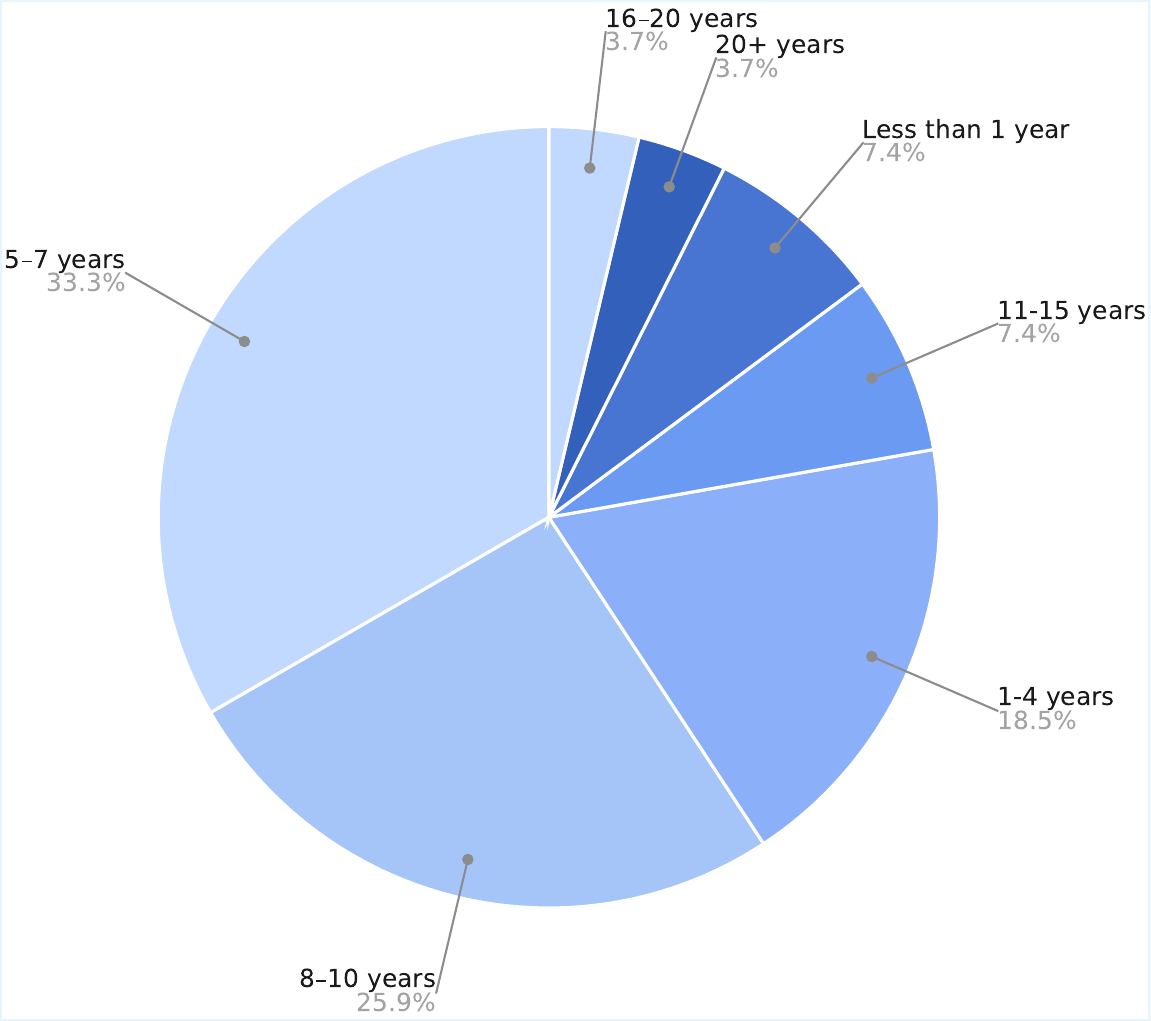}
            \caption{Machine Learning Development}
            \label{fig:ml_experience}
        \end{subfigure}
        
        \caption{Distribution of survey respondents’ professional experience across (a) software engineering, (b) Python programming, and (c) machine learning development.}
        \label{fig:experience_distribution}
    \end{figure*}

\begin{figure}[t]
    \centering
    \includegraphics[width=\textwidth]{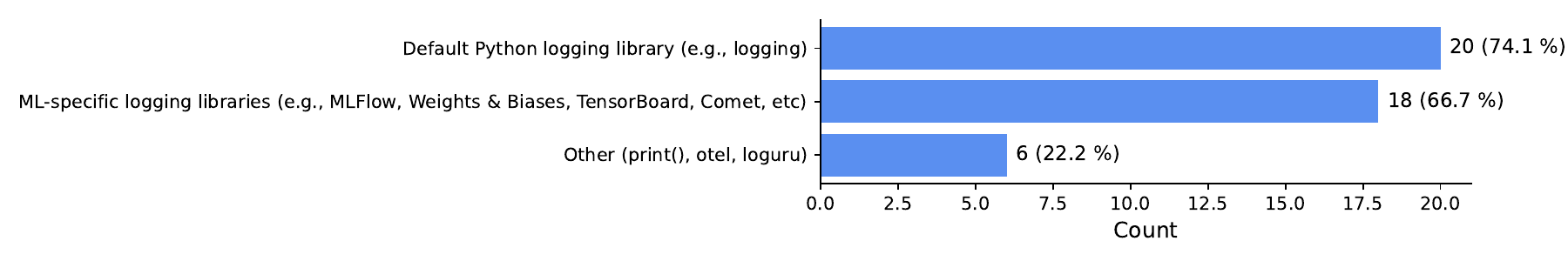}
    \caption{Logging libraries used by survey respondents in ML projects.}
    \label{fig:logging_libraries}
\end{figure}

\begin{figure}[t]
    \centering
    \includegraphics[width=\textwidth]{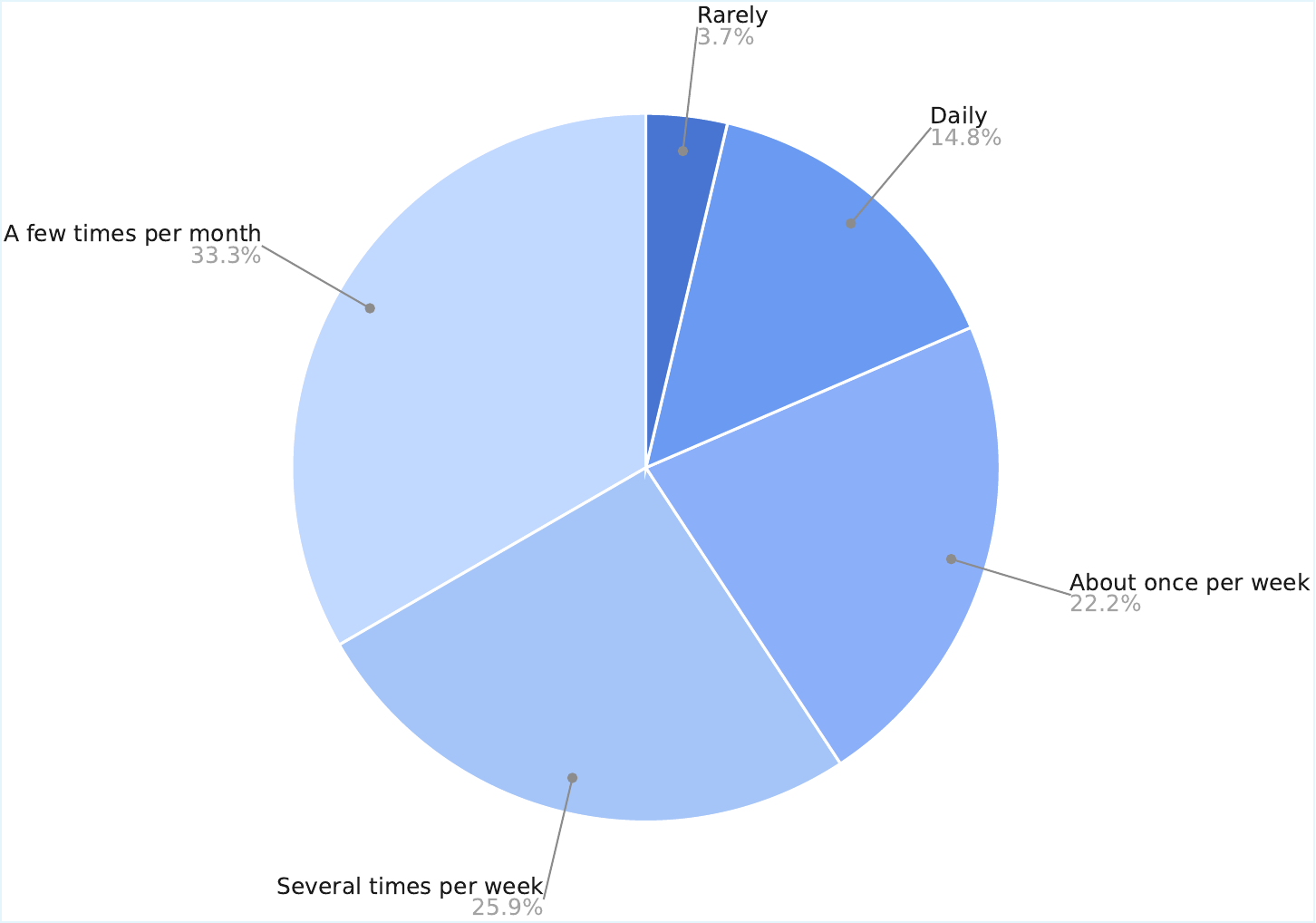}
    \caption{Frequency with which respondents add or modify logging statements in their projects.}
    \label{fig:modify_logging_frequency}
\end{figure}

We received 27 complete responses to the survey. All participants accepted to participate and reported their professional roles and backgrounds.\\
    Figure~\ref{fig:participant_roles} presents the distribution of respondents’ professional roles. The sample is dominated by practitioners with strong ML and research-oriented profiles: Research Scientists/Applied Scientists represent 29.6\% (8/27) of respondents, followed by ML Engineers at 14.8\% (4/27). Software Developers/Engineers and PhD Students each account for 11.1\% (3/27). Data Scientists, Senior Software Engineers, and DevOps/SRE/Platform Engineers each represent 7.4\% (2/27), while Engineering Managers, Tech Leads/Team Leads, and Staff/Principal Engineers each account for 3.7\% (1/27). This distribution indicates that our respondents are primarily hands-on ML practitioners. This profile is reinforced by Figure~\ref{fig:experience_distribution}, which demonstrates significant professional experience across software engineering, Python programming, and machine learning development. In software engineering, 70.3\% of respondents report 5–15 years of experience, and 7.4\% report more than 20 years. In Python programming, 70.3\% have between 5 and 10 years of experience, and in machine learning development, 66.6\% report 5–15 years. Overall, the sample reflects considerable exposure to real-world ML systems, lending credibility to their assessments of logging practices.\\
Regarding tooling, Figure~\ref{fig:logging_libraries} indicates that 66.7\% of respondents use ML-specific logging libraries (e.g., MLflow, Weights \& Biases, TensorBoard, Comet), while 74.1\% also rely on the default Python \texttt{logging} library. Additionally, 22.2\% reported using alternative mechanisms such as \texttt{print()}, \texttt{otel}, or \texttt{loguru} through open-ended responses. These observations highlight the hybrid logging landscape in ML systems, where practitioners combine general-purpose logging with ML-specific tracking frameworks.
Figure~\ref{fig:modify_logging_frequency} further shows that logging is actively maintained by practitioners: 25.9\% modify logging statements several times per week, 14.8\% do so daily, 22.2\% about once per week, only 3.7\% rarely, and the remainder at least monthly. This confirms that logging is a continuous engineering activity in ML workflows, underscoring the practical importance of identifying and mitigating logging smells.
Finally, we assess practitioners’ opinions on the taxonomy along 4 complementary dimensions—perception of the smell taxonomy, frequency, relevance, and severity.

\paragraph{\textbf{Perception of the Smell Taxonomy}}

\begin{figure}[t]
    \centering
    \includegraphics[width=\textwidth]{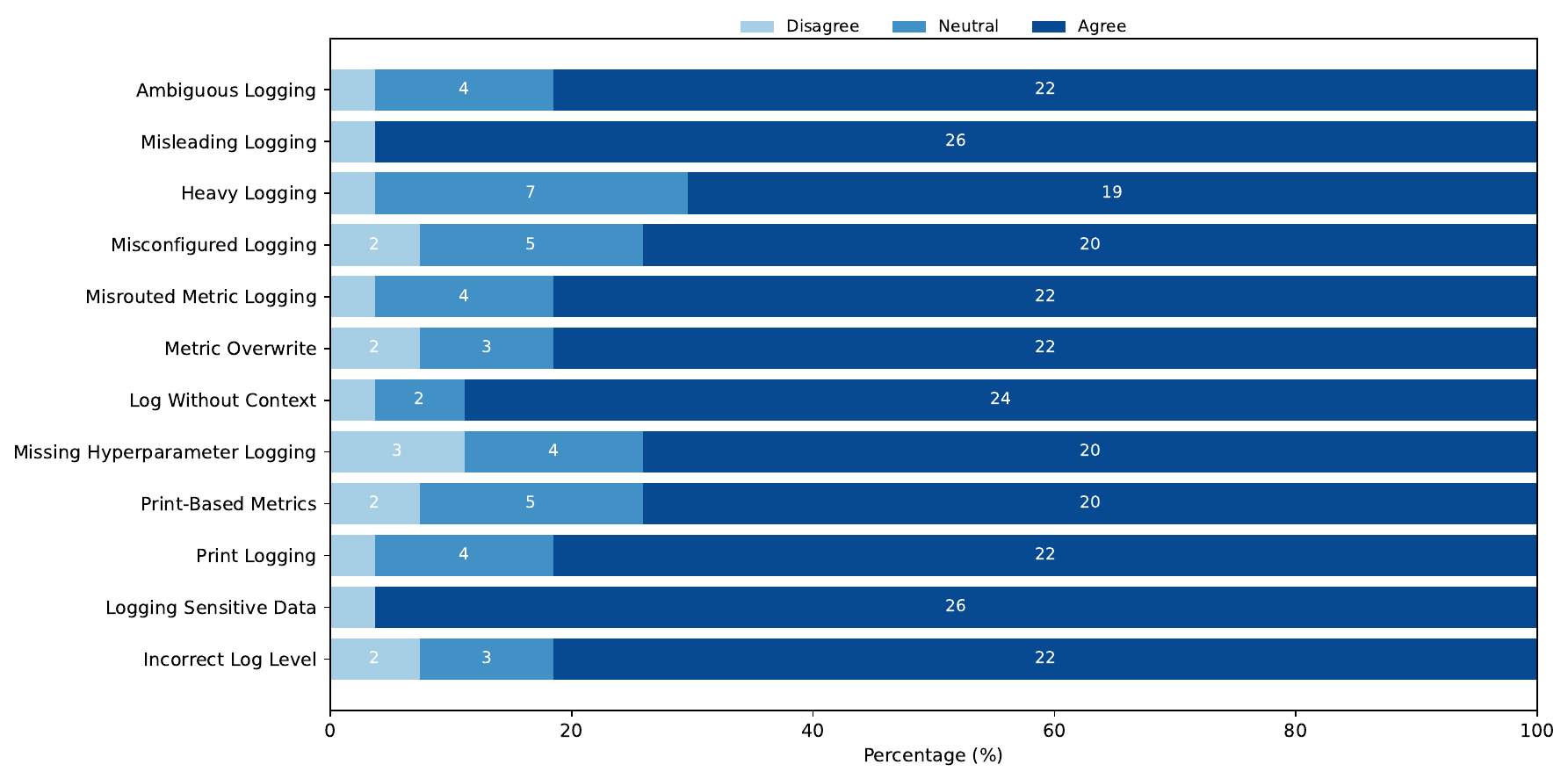}
    \caption{Distribution of survey responses on agreement with each identified logging smell.}
    \label{fig:agreement_smells}
\end{figure}

Figure~\ref{fig:agreement_smells} highlights an important observation regarding the validity of the proposed logging smell taxonomy. For most smells, the majority of respondents selected ``Agree,'' indicating broad recognition of these issues as genuine problems in ML systems. In particular, Misleading Logging, Logging Sensitive Data, Missing Hyperparameter Logging, Metric Overwrite, Misleading Logging, and Log Without Context received near-unanimous agreement.

Nevertheless, a small but meaningful proportion of neutral or disagree responses emerged for certain smells, notably \textit{Heavy Data Logging, Ambiguous Logging, Misconfigured Logging, Print-Based Metrics, and Print Logging}. The open-ended question—\textit{``In your own words, what makes this logging smell problematic in practice? Please describe any specific technical issues, failures, or challenges you have personally encountered when dealing with this smell.''}—provided qualitative insights that help contextualize these divergences.

For \emph{Ambiguous Logging}, practitioners emphasized issues such as unclear metric names, insufficient contextual information, and misunderstandings within teams. Some respondents noted that ambiguous metric labels lose meaning over time or are only interpretable by their original authors. Others argued that ambiguity is highly context-dependent or overlaps conceptually with other smells (e.g., misleading logging), suggesting the need for clearer conceptual boundaries within the taxonomy.

\begin{center}
    \includegraphics[width=2ex]{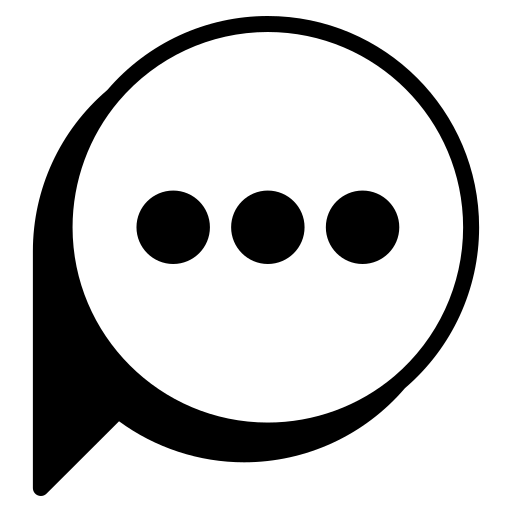} \textit{``I think the major issue is the lack of details on the metrics. If you have many metrics to log, this will cause a problem.''}

    \includegraphics[width=2ex]{chat-box.png} \textit{``Ambiguous logging names quickly lose their meaning, or are only meaningful to those who introduced them. I don't agree with mixing this smell with contradictory names (train vs. valid); this is outright misleading and should be a different category.''}

    \includegraphics[width=2ex]{chat-box.png} \textit{``In principle, ambiguity seems like the biggest potential problem, particularly with large or mixed development teams. For a metric as ambiguous as `x', the logging is practically worthless or even counterproductive, although there is a spectrum of ambiguity.''}
\end{center}

For \emph{Heavy Data Logging}, disagreement often reflected perceived trade-offs rather than rejection of the concept itself. Several practitioners viewed extensive logging as useful during experimentation or system debugging, arguing that performance overhead is situational and often manageable. Thus, heavy data logging was sometimes framed as a performance optimization concern rather than an intrinsic design flaw, revealing tensions between observability and efficiency.

\begin{center}
    \includegraphics[width=2ex]{chat-box.png} \textit{``It's good to do this when figuring out the system.''}

    \includegraphics[width=2ex]{chat-box.png} \textit{``Computing additional statistics as in the example does not seem that problematic in terms of overhead. Doing additional inference steps could possibly be done in a faster way outside of the logging function, but this depends on the framework. I don't see why it should be discouraged in general.''}

    \includegraphics[width=2ex]{chat-box.png} \textit{``I believe this case brings too much redundant information to log, which may increase computational burden and make analysis harder afterwards.''}
\end{center}

Similarly, \emph{Misconfigured Logging} elicited neutral responses when respondents considered small-scale projects or modern ML frameworks (e.g., experiment-tracking tools) that abstract configuration management. Some participants noted that centralized configuration is context-dependent and may not always introduce practical harm, particularly in smaller or well-structured projects.

\begin{center}
    \includegraphics[width=2ex]{chat-box.png} \textit{``Sure, it's great to have centralized logging for a large, maintained project, but for smaller projects I sometimes define it at a single point where it is convenient.''}

    \includegraphics[width=2ex]{chat-box.png} \textit{``In ML settings it is sometimes necessary to have this. Frameworks like wandb handle this correctly.''}

    \includegraphics[width=2ex]{chat-box.png} \textit{``I can see that this is problematic when reusing code in different contexts if the logging configuration is tightly coupled to the reused code (e.g., inside a class constructor). In a launch script, however, I can also see advantages in not having to manage separate configuration files.''}
\end{center}

The strongest divergence appeared for \emph{Print-Based Metrics} and \emph{Print Logging}. While many respondents acknowledged their limitations—particularly in infrastructure code, large-scale debugging, or log aggregation—others considered them acceptable in controlled or experimental settings. Some practitioners reported redirecting standard output to files or using command-line tools (e.g., \texttt{grep}, \texttt{awk}) to filter output, thereby mitigating the perceived severity. In ML experimentation contexts, print-based output was sometimes viewed as pragmatic and lightweight rather than inherently problematic.

\begin{center}
    \includegraphics[width=2ex]{chat-box.png} \textit{``I think for infrastructure code, this is definitely an issue. For some ML debugging, however, this happens.''}

    \includegraphics[width=2ex]{chat-box.png} \textit{``For me this is not severe at all. Once you understand the logging structure and get used to it, it is not hard to filter output using \texttt{grep}, \texttt{awk}, or similar tools.''}

    \includegraphics[width=2ex]{chat-box.png} \textit{``Based on my programming style, I redirect standard output to a specific file at the beginning of each process, making print statements almost equivalent to logging. However, for standard software engineering, this can be critical because stdout has a specific meaning.''}

    \includegraphics[width=2ex]{chat-box.png} \textit{``It loses the log level significance.''}

    \includegraphics[width=2ex]{chat-box.png} \textit{``Unlike the previous one, this can create challenges for system debugging by making it harder to identify the root cause in system logs.''}
\end{center}

Overall, the agreement distribution, combined with qualitative feedback, indicates that the taxonomy is largely aligned with practitioners’ lived experience. At the same time, it reveals nuanced, context-sensitive interpretations—particularly for smells related to verbosity, configuration practices, and print-based logging. This balance between strong consensus and constructive boundary discussions strengthens the empirical grounding and practical relevance of the proposed taxonomy.

\paragraph{\textbf{Perception of the Smell's Frequency}}

\begin{figure}[t]
    \centering
    \includegraphics[width=\textwidth]{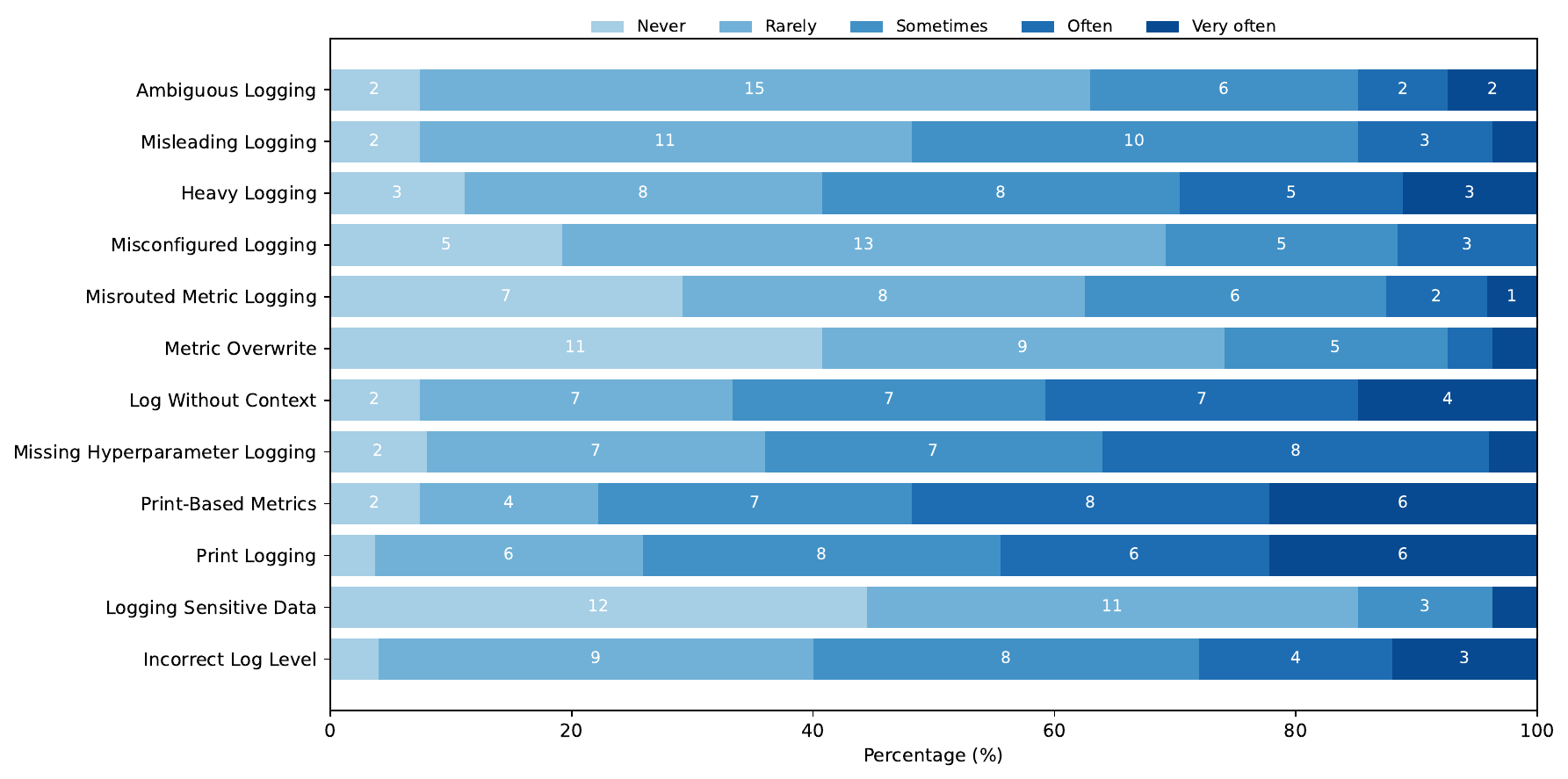}
    \caption{Distribution of survey responses on the frequency of encountering each logging smell.}
    \label{fig:frequency_smells}
\end{figure}

Figure~\ref{fig:frequency_smells} presents practitioners’ responses to the question, ``How frequently have you observed this smell in your projects?'' Overall, the results indicate that several logging smells are not perceived as rare anomalies but rather as recurring issues in ML development. In particular, Print Logging, Print-Based Metrics, Log Without Context, Missing Hyperparameter Logging, and Heavy Data Logging exhibit high proportions of responses in the ``Often'' and ``Very often'' categories, suggesting that these practices are commonly encountered in real-world systems. 

By contrast, smells such as Logging Sensitive Data and Metric Overwrite tend to concentrate in the ``Rarely'' or ``Sometimes'' categories. Ambiguous Logging and Misleading Logging display a mixed distribution, with many respondents selecting ``Rarely'' or ``Sometimes,'' but a non-negligible proportion reporting encountering them ``Often,'' highlighting variability across teams and projects.

Importantly, very few smells are predominantly rated as ``Never,'' indicating that most of the identified issues have been experienced at least occasionally by practitioners. Taken together, these observations suggest that the proposed smells capture recurrent patterns in ML logging practice, with some representing pervasive day-to-day concerns (e.g., verbosity and incomplete context) and others representing less frequent but still significant risks. This frequency distribution reinforces the practical relevance of the taxonomy and supports the need for systematic guidance and tooling support to mitigate these recurring logging issues in ML systems.

\paragraph{\textbf{Perception of the Smell's Relevance}}

\begin{figure}[t]
    \centering
    \includegraphics[width=\textwidth]{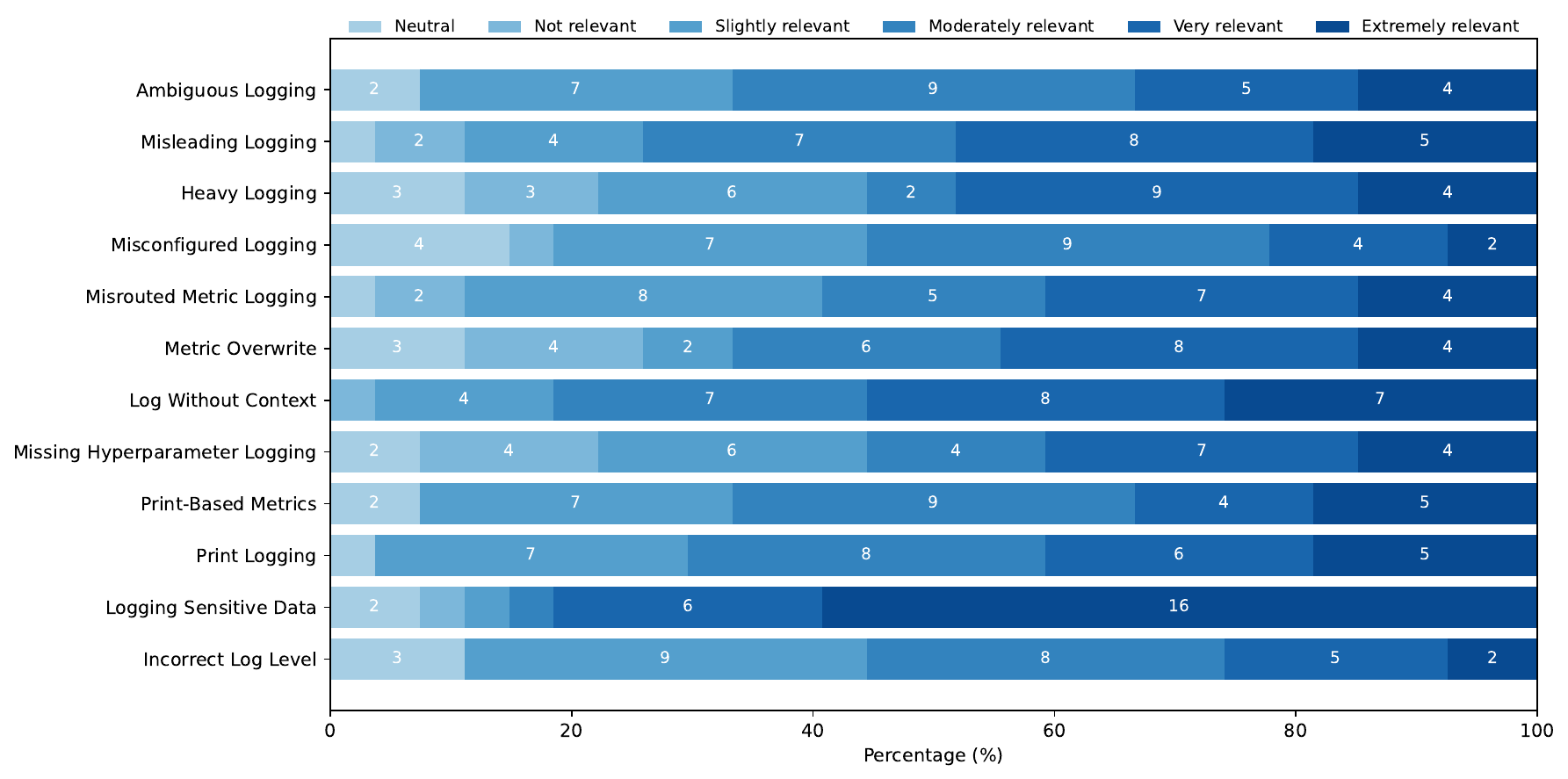}
    \caption{Distribution of survey responses on the perceived relevance of addressing each logging smell.}
    \label{fig:relevance_smells}
\end{figure}

Figure~\ref{fig:relevance_smells} reports practitioners’ answers to: ``How relevant is it to address this smell in practice?'' Overall, respondents perceived most smells as actionable and worth addressing, with responses generally concentrated in the \emph{moderately} to \emph{extremely relevant} range. Relevance was highest for smells that directly threaten security, correctness, and reproducibility. For instance, \emph{Logging Sensitive Data} was rated \emph{very} or \emph{extremely relevant} by 22/27 respondents (81.48\%), and open-ended comments consistently framed it as a security and compliance risk (e.g., ``security risk \ldots law compliance risk''; ``Don't log your API keys''). Similarly, \emph{Misleading Logging}, \emph{Log Without Context}, and \emph{Missing Hyperparameter Logging} were widely judged relevant, with qualitative responses emphasizing that these issues impair debugging, traceability, and experiment reproducibility (e.g., missing context makes logs ``practically worthless'' for diagnosis, while missing hyperparameters undermines reliable comparison and replication of runs). In contrast, relevance perceptions were more mixed for smells whose impact depends on project scale, tooling, or performance constraints. \emph{Misconfigured Logging} attracted the largest share of low-relevance responses (5/27 neutral or not relevant), as some practitioners argued that configuration choices are acceptable in small projects or are handled by modern ML tooling (e.g., experiment-tracking frameworks), while others highlighted maintainability issues when code is reused across contexts. Likewise, \emph{Heavy Data Logging} and \emph{Metric Overwrite} received a small but notable minority of ``not relevant'' assessments: respondents described heavy data logging as sometimes beneficial during experimentation (with overhead ``situational and manageable''), and metric overwrite as either rare or easy to detect, despite others warning that it can silently misrepresent progress and results. Taken together, these observations suggest that practitioners view the taxonomy as practically relevant, while also highlighting a subset of smells whose perceived urgency is shaped by context (e.g., debugging phase, performance sensitivity, and the surrounding logging/tooling ecosystem).

\paragraph{\textbf{Perception of the Smell's Severity}}

\begin{figure}[t]
    \centering
    \includegraphics[width=\textwidth]{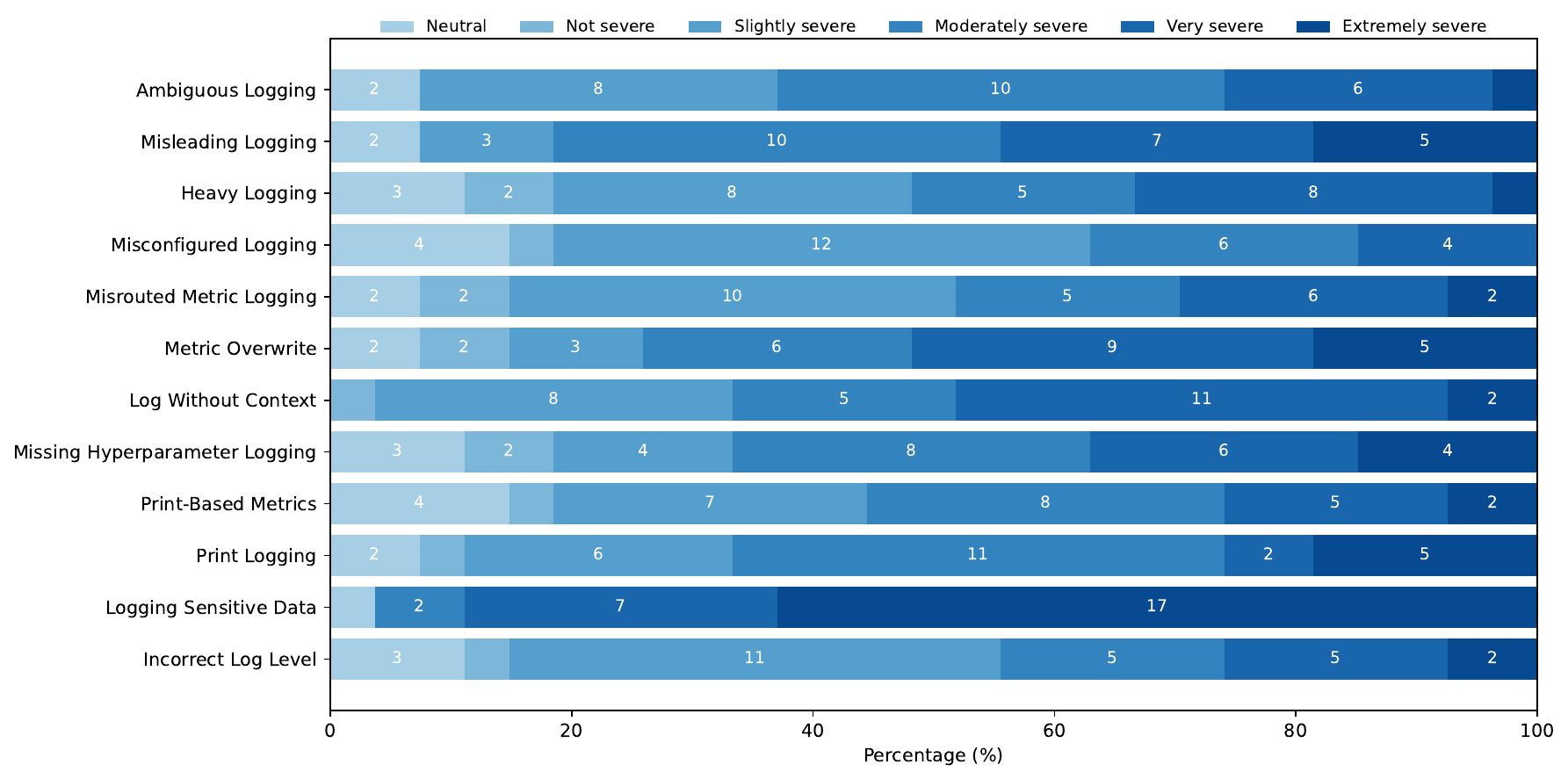}
    \caption{Distribution of survey responses on the perceived severity of each logging smell.}
    \label{fig:severity_smells}
\end{figure}

Figure~\ref{fig:severity_smells} presents practitioners’ assessments of the severity of each logging smell. Overall, several smells are perceived as not only relevant but also highly severe, particularly those that directly affect security, correctness, and reproducibility. \emph{Logging Sensitive Data} clearly stands out as the most severe smell, with the vast majority of respondents rating it as \emph{very} or \emph{extremely severe}. Open-ended responses consistently frame this issue as a critical security and compliance risk, especially in production or regulated environments.

\begin{center}
    \includegraphics[width=2ex]{chat-box.png} \textit{``Security risk \ldots law compliance risk.''}

    \includegraphics[width=2ex]{chat-box.png} \textit{``Don't log your API keys.''}

    \includegraphics[width=2ex]{chat-box.png} \textit{``An extremely serious cybersecurity problem — deciding what to log and what not to log is at the very core of logging.''}
\end{center}

Similarly, \emph{Metric Overwrite}, \emph{Misleading Logging}, \emph{Log Without Context}, and \emph{Missing Hyperparameter Logging} are predominantly rated as \emph{moderately} to \emph{very severe}. Practitioners emphasized that these smells can silently distort experimental results, hinder debugging, and undermine reproducibility—core requirements in ML systems.

\begin{center}
    \includegraphics[width=2ex]{chat-box.png} \textit{``Overwritten metrics give a false impression that `it's working', \ldots completely misrepresenting the real information.''}

    \includegraphics[width=2ex]{chat-box.png} \textit{``It could ruin a costly training run.''}

    \includegraphics[width=2ex]{chat-box.png} \textit{``Reduces reproducibility.''}

    \includegraphics[width=2ex]{chat-box.png} \textit{``If context is missing and a problem occurs, reading the logs won't be helpful at all.''}
\end{center}

\emph{Misconfigured Logging} exhibits more heterogeneous severity assessments, ranging from minor to severe. Several practitioners described it as context-dependent, particularly in small projects or when modern ML frameworks abstract configuration management. However, others emphasized structural risks, especially when logging configuration is tightly coupled to reusable components or globally overridden by libraries.

\begin{center}
    \includegraphics[width=2ex]{chat-box.png} \textit{``A library should never attempt to globally configure logging that overrides the user's configuration.''}

    \includegraphics[width=2ex]{chat-box.png} \textit{``Hard to identify configuration details during debugging.''}

    \includegraphics[width=2ex]{chat-box.png} \textit{``Goes against the concept of core libraries \ldots everyone must use the same piece.''}
\end{center}

These responses suggest that misconfiguration can compromise maintainability and composability, particularly in shared or large-scale codebases, even if it is sometimes perceived as harmless in controlled contexts.

\emph{Misrouted Metric Logging} is generally perceived as moderately to highly severe, particularly when it leads to incomplete reporting, inconsistent experiment tracking, or lost metrics. Although some respondents noted that such issues may be quickly detected and corrected, others stressed the potential for wasted computation and misleading dashboards.

\begin{center}
    \includegraphics[width=2ex]{chat-box.png} \textit{``It strips the ability to do statistical analysis and comparisons between runs.''}

    \includegraphics[width=2ex]{chat-box.png} \textit{``Makes reporting incomplete.''}

    \includegraphics[width=2ex]{chat-box.png} \textit{``You don’t fully realize which metrics should be logged \ldots after training you don’t have the necessary ones, leading to retraining.''}
\end{center}

Some practitioners characterized it as a straightforward coding error:

\begin{center}
    \includegraphics[width=2ex]{chat-box.png} \textit{``If you need the metric you'll notice and fix it.''}

    \includegraphics[width=2ex]{chat-box.png} \textit{``I treat this as a bug.''}
\end{center}

However, even when fixable, its consequences may be costly if discovered late in long-running experiments or multi-process training setups.

In contrast, \emph{Heavy Data Logging}, \emph{Print-Based Metrics}, \emph{Print Logging}, and \emph{Incorrect Log Level} exhibit greater variability in perceived severity. While a substantial proportion of respondents rate them as \emph{very severe}, a non-negligible share classify them as \emph{slightly severe} or \emph{minor}. Qualitative responses indicate that these smells are often viewed as trade-offs between observability, performance, and developer convenience.

\begin{center}
    \includegraphics[width=2ex]{chat-box.png} \textit{``It's good to do this when figuring out the system.''}

    \includegraphics[width=2ex]{chat-box.png} \textit{``It really depends on the context. In production code this is a bad idea.''}

    \includegraphics[width=2ex]{chat-box.png} \textit{``It may not break the system, but using an incorrect log level can hide important information during debugging.''}
\end{center}

Overall, the severity distribution indicates that practitioners clearly distinguish between smells that pose systemic risks—such as security breaches, corrupted metrics, loss of consistency, and loss of reproducibility—and those that primarily introduce performance, readability, or maintainability concerns. The inclusion of \emph{Misconfigured Logging} and \emph{Misrouted Metric Logging} further highlights that configuration and metric consistency are critical in ML systems, where experiment tracking, comparability, and large-scale execution amplify the consequences of seemingly small logging design flaws. This nuanced perception reinforces the taxonomy’s empirical validity while emphasizing that severity is shaped by project context, system scale, and the surrounding tooling ecosystem.

    \begin{tcolorbox}[colback=black!4,colframe=black!50!white]
    ML practitioners largely recognize the identified logging smells as real and practically meaningful issues, with strong agreement and high perceived relevance and severity for smells that threaten security, reproducibility, and correctness (e.g., \textit{Logging Sensitive Data, Metric Overwrite, Missing Hyperparameter Logging}). Smells related to verbosity, configuration, and print-based practices are perceived more contextually, with their severity and relevance shaped by project scale, tooling ecosystem, and development phase.
    \end{tcolorbox}

\section{Implications}
\label{sec:discussion_implication}
In this section, we discuss the implications of our findings for researchers, practitioners, and educators.
\subsection{Implications for Researchers}
Our findings confirm that logging smells in ML systems constitute a distinct and underexplored research area. Practitioners not only recognize these smells as real problems but also describe additional dimensions that extend beyond our taxonomy, including interleaved logging across experiments, logging frequency imbalance (too often or too rarely) in ML system, logging overhead in GPU-intensive workflows, structured logging misuse, missing logging in critical code paths, and conflicts between multiple logging libraries.

For example, one practitioner reported:
\begin{center}
\includegraphics[width=2ex]{chat-box.png} \textit{``Interleaved logging; logging metrics from multiple experiments melding into the same metric or file \ldots producing alternating metrics from the involved experiments.''}

\includegraphics[width=2ex]{chat-box.png} \textit{``Maybe logging ML metrics too often, or too rarely, or logging too much (e.g., PyTorch Lightning makes it easy to log gradient magnitudes with wandb, but doing so unconditionally creates a lot of data that typically nobody looks at).''}
\end{center}

Another emphasized performance implications:
\begin{center}
\includegraphics[width=2ex]{chat-box.png} \textit{``Printing / logging the value of some objects takes time \ldots a CUDA tensor norm incurs a CUDA sync. Many disregard that and just pay the price.''}
\end{center}

Importantly, to foster future research, we publicly release a labeled dataset of 2,448 logging smell instances derived from ML projects. This dataset enables the research community to extend our work in multiple directions. First, it provides a foundation for developing automated logging smell detection techniques using AI and large language models (LLMs). Second, it supports research on automated repair or refactoring of logging smells, where AI systems suggest or implement corrective transformations. Notably, practitioners themselves alluded to the growing role of LLMs in improving logging practices—both positively and negatively. While some respondents observed that LLMs can improve experiment tracking and produce structured logging patterns,

\begin{center}
\includegraphics[width=2ex]{chat-box.png} \textit{``LLM generated code is fixing some of this as they code good experiment tracking code.''}

Others reported that LLM-generated logging may introduce new forms of logging smells, particularly excessive or redundant logging:

\includegraphics[width=2ex]{chat-box.png} \textit{``LLMs tend to log EVERYTHING and it's a real pain, you end up with 120 different semantically overlapping logs, it's a disaster''}

\includegraphics[width=2ex]{chat-box.png} \textit{``LLM tend to do these things, it's painful. ''}
\end{center}

This observation suggests a promising research avenue at the intersection of logging quality and AI-assisted development, including (i) training models to detect logging smells, (ii) generating context-aware logging fixes, and (iii) evaluating whether AI-generated logging adheres to reproducibility and observability best practices or introduces specific logging smells.

Beyond training-time logging, practitioners also highlighted challenges specific to inference environments and complex operators:

\begin{center}
\includegraphics[width=2ex]{chat-box.png} \textit{``This may be specific to inference engines, but for some complex operators (ex. Attention, PagedAttention), I would love to have standardized logging - if there was a way to know at which intermediate tensors I need to pay particular attention to, it would be perfect..''}
\end{center}

This suggests a need to investigate logging smells in inference pipelines, where latency constraints, distributed execution, and observability requirements differ significantly from training workflows.

Finally, while our study focuses on Python-based ML systems, logging smells are not language-bound. Several practitioners referred to practices inherited from imperative programming styles (e.g., C/C++, Bash), and issues such as logger collisions, global configuration overrides, and structured logging misuse are equally relevant in ecosystems such as Java, C/C++, and large-scale production systems. Future work should therefore examine logging smells across programming languages and ML frameworks to assess cross-language generalizability, tooling differences, and ecosystem-specific manifestations.

Overall, our results, combined with the released dataset and practitioner insights, position logging smell research as a fertile area for advancing automated quality assurance, AI-assisted repair, cross-language empirical studies, and inference-stage observability engineering in ML systems.

\subsection{Implications for Educators}

The survey responses reveal that logging in ML systems is rarely treated as a first-class engineering topic. One participant noted:
\begin{center}
\includegraphics[width=2ex]{chat-box.png} \textit{``Many ML researchers don't think of these as actual issues but things that they get used to because the codebase allows it.''}
\end{center}

This observation indicates a structural gap in ML education, where experiment tracking, observability, and logging design are often secondary to model accuracy and algorithmic performance. To better understand the educational landscape, we conducted a targeted search of publicly available lecture notes and course materials on machine learning systems, trustworthy machine learning, and observability in ML systems using Google search \footnote{\url{https://www.google.com/}}. We identified courses from major institutions—including the University of Wisconsin–Madison, Massachusetts Institute of Technology, University of Washington, University of Toronto, Stanford University, and the University of Notre Dame. While these courses comprehensively cover ML algorithms, fairness, ethics, and system scalability, very few explicitly address experiment tracking, logging design, or logging quality in ML systems \citep{Stanford}. 

This gap suggests that logging and observability are often implicitly assumed rather than explicitly taught. As ML systems become increasingly complex and deployed in high-stakes environments, experiment traceability, reproducibility, and logging correctness should be integrated into core ML systems curricula. Educators should therefore incorporate modules on structured logging, metric lifecycle management, experiment tracking frameworks, security-aware logging, configuration management, and logging performance trade-offs.

Moreover, feedback on survey clarity suggests that terminology such as ``logging smells'' is not universally familiar among practitioners. This highlights the need to formally define logging smell in educational materials and to provide.

Overall, our paper can serve as a foundation for introducing dedicated lectures or modules on logging and logging smells in ML systems, targeting both ML researchers and practitioners. Framing logging as a core engineering discipline—at the intersection of software architecture, security, performance engineering, and data governance—can significantly improve the reliability, reproducibility, and trustworthiness of ML-based applications.

\section{Threats to validity}
\label{sec:threats_to_validity}
This section discusses potential threats to the validity of the study and the strategies we adopt to mitigate them.

\textit{Internal Validity.}  
A primary internal threat concerns potential confounding factors that may affect the accuracy of our analyses. For RQ1, the automated extraction and LLM-based classification of logging statements may introduce false positives or false negatives, thereby distorting the actual distribution of logging smells. To reduce this risk, we employed a human-in-the-loop validation process: multiple authors independently reviewed LLM outputs--either jointly during calibration meetings or separately--and discrepancies were resolved through discussion. We computed Cohen’s kappa to quantify inter-rater reliability, obtaining a score of 0.84, which indicates strong agreement among reviewers~\citep{cohen1960coefficient}.

For RQ2, several threats stem from the survey methodology. First, participant self-selection bias may arise if individuals with strong opinions about logging or greater ML experience are more motivated to respond. To mitigate this, we recruited participants from diverse sources, including contributors to the 444 open-source ML projects examined in our study as well as practitioners from industry communities. This strategy ensured that respondents had direct experience with ML development and increased the relevance of the collected insights. We also incorporated background questions into the survey to assess respondents’ roles, years of experience, and familiarity with ML logging practices, allowing us to verify that participants possessed sufficient domain knowledge and enabling us to contextualize their responses appropriately. Finally, we report respondent demographics to promote transparency and strengthen the interpretability of the results.

We offered a \$100 Amazon gift card raffle to encourage participation, which helped broaden the respondent pool but may also have introduced sampling bias by attracting individuals primarily motivated by the incentive. Additionally, the survey contained approximately 36 questions, estimated to take 20-30 minutes to complete. Longer surveys may reduce respondent attention and lead to satisficing behaviors or random answers, particularly after ten minutes of engagement~\citep{kost2018impact}. This factor may affect the precision of certain responses.

Another threat concerns the potential misunderstanding of smell definitions by participants. To alleviate this, we provided each smell with a clear description and a representative example. Furthermore, we conducted a pilot study with 5 participants to assess clarity and comprehensibility. Based on their feedback, we refined the survey instrument to ensure that the questions and smell descriptions were accessible and unambiguous.

\textit{Construct Validity.}  
Threats to construct validity concern the degree to which our operationalizations and measurements accurately capture the phenomena under study. For RQ1, the process of identifying and labeling ``logging smells'' may be subject to interpretive bias during manual annotation. To reduce this risk, we adopted a structured open-coding procedure augmented by LLM-assisted suggestions, followed by independent verification by two researchers to ensure consistent interpretation of smell categories.

Another potential threat arises from inaccuracies in extracting functions containing logging statements, particularly the risk of inadvertently including commented-out code. Such cases could distort the true prevalence and nature of logging practices. To address this, we developed a custom static code analysis tool built on top of Python’s Abstract Syntax Tree (AST) infrastructure--a commonly used mechanism for Python static analysis~\citep{foalem2024studying, dilhara2021understanding}. Our analyzer, which is included in the replication package~\citep{replication}, specifically filters out commented functions and extracts only active (uncommented) function definitions. This approach minimizes noise in the dataset and ensures that all analyzed logging statements reflect developer-intended behavior.

\textit{External Validity.}  
A primary external threat to validity concerns the representativeness of our dataset and the extent to which our findings generalize beyond the analyzed sample. Our dataset, consisting of 444 recent and active open-source Python-based ML projects, offered a strong empirical foundation but may still limit generalizability to other programming languages or proprietary industrial systems. Likewise, our analysis relied on a predefined set of logging libraries identified in prior work and drawn from a dataset released within the last two years. Although this ensured that our sample reflected contemporary ML development practices, it may not fully capture emerging or domain-specific logging frameworks introduced more recently.
To mitigate these limitations, we complemented the repository mining with a practitioner survey (RQ2), enabling us to incorporate perspectives from contributors to open-source projects as well as practitioners working in industrial environments. This methodological triangulation strengthened the external validity of our findings by grounding them in both observed code and practitioner experience.
As a potential avenue for future work, the dataset could be expanded to include projects written in other programming languages (e.g., Java, C++, R) or to track newer logging libraries and ML observability tools as they evolve, thereby enabling a broader assessment of whether the identified logging smells persist across different technological ecosystems and development contexts.

\section{Conclusion and future work}
\label{sec:conclusion}
Logging plays a central role in ensuring reproducibility, observability, and reliability in machine learning (ML) systems. In this study, through mining open-source ML projects and qualitative analysis, we identified a taxonomy of 12 logging smell categories spanning both general-purpose and ML-specific concerns, including security, experiment tracking, metric management, configuration, and verbosity. The relevance and severity of these smells are further validated through a survey of 27 ML practitioners, confirming that several categories—such as \textit{Logging Sensitive Data, Metric Overwrite, Missing Hyperparameter Logging,} and \textit{Log Without Context}—are particularly critical in practice.

This study has several implications. First, researchers can leverage our publicly released dataset to build automated logging smell detection models using AI and large language models (LLMs). Second, our taxonomy provides a foundation for developing automated repair techniques to refactor or correct problematic logging patterns. Third, practitioners can incorporate these smell categories into logging guidelines and experiment management practices. Finally, educators can integrate logging quality and observability as core topics in ML systems curricula.

In future work, we plan to develop automated approaches for detecting and repairing logging smells using machine learning and LLM-based techniques. We also intend to investigate logging smells in inference pipelines, where latency and deployment constraints introduce new challenges. Another important direction is to study logging smells across different programming languages (e.g., C/C++, Java) and ML ecosystems to assess cross-language generalizability. By advancing automated detection, repair, and cross-ecosystem studies, we aim to strengthen the reliability and trustworthiness of ML-based software systems.

\section*{Declarations}
\label{sec:declaration}
\textbf{Conflicts of interests/Competing interests:} The authors declare that they have no known competing financial interests or personal relationships that could have appeared to influence the work reported in this paper.

\noindent\textbf{Data availability statement:} The datasets generated during and/or analysed during the current study are available in the [foalem] repository, [\url{https://github.com/foalem/LoggingSmellMLCode}].

\noindent\textbf{Ethics Declaration} This study was approved by Ethics Committee of Polytechnique Montreal (CER-2324-25-D). All participants provided informed consent prior to their participation.

\noindent\textbf{Informed Consent} All participants provided informed consent before taking part in the survey.












\bibliographystyle{plainnat}
\bibliography{Paper}

\end{document}